\documentclass[aps,prd,twocolumn,superscriptaddress,nofootinbib]{revtex4}

\usepackage{hyperref}
\usepackage{amsmath,amssymb,amsfonts}
\usepackage{color}
\usepackage{graphicx}
\usepackage{enumerate} 
\usepackage{colordvi} 
\usepackage{bm}
\usepackage{multirow}

\newcommand{\be}{\begin{equation}}
\newcommand{\ee}{\end{equation}}
\newcommand{\nn}{\nonumber}

\newcommand{\bfk}{\boldsymbol{k}}
\newcommand{\bfr}{\boldsymbol{r}}
\newcommand{\bfv}{\boldsymbol{v}}
\newcommand{\bfs}{\boldsymbol{s}}
\newcommand{\bfx}{\boldsymbol{x}}
\newcommand{\bfz}{\boldsymbol{z}}

\newcommand{\Omegam}{\Omega_{\rm m}}
\newcommand{\OmegaK}{\Omega_K}
\newcommand{\OmegaDE}{\Omega_{\rm DE}}

\newcommand{\wt}{\widetilde}

\newcommand{\PEE}{P_{EE}}
\newcommand{\Pgg}{P_{gg}}

\newcommand{\PgE}{P_{gE}}

\newcommand{\sigv}{\sigma_{\rm v}}
\newcommand{\sigvlin}{\sigma_{\rm v,lin}}
\newcommand{\Dfog}{D_{\rm FoG}}

\newcommand{\Pdd}{P_{\delta\delta}}
\newcommand{\Pdt}{P_{\delta\theta}}
\newcommand{\Ptt}{P_{\theta\theta}}
\newcommand{\Pdg}{P_{\delta g}}

\newcommand{\xiddl}{\xi_{\delta\delta,\ell}}

\newcommand{\xidd}{\xi_{\delta\delta}}

\newcommand{\xidg}{\xi_{\delta g}}
\newcommand{\xigg}{\xi_{gg}}

\newcommand{\deltags}{\delta_{g}^S}
\newcommand{\deltag}{\delta_{g}}
\newcommand{\gammaijs}{\gamma_{ij}^S}

\newcommand{\xigps}{\xi_{g+}^S}

\newcommand{\xims}{\xi_{-}^S}

\newcommand{\xiXs}{\xi_{X}^S}

\newcommand{\PhgEs}{\hat{P}_{gE}^S}
\newcommand{\PhEEs}{\hat{P}_{EE}^S}
\newcommand{\PhgBs}{\hat{P}_{gB}^S}
\newcommand{\PhBBs}{\hat{P}_{BB}^S}
\newcommand{\Phggs}{\hat{P}_{gg}^S}
\newcommand{\Phgps}{\hat{P}_{g+}^S}
\newcommand{\Phgms}{\hat{P}_{g\times}^S}
\newcommand{\Phppmms}{\hat{P}_{+\times}^S}
\newcommand{\Phmmpps}{\hat{P}_{\times+}^S}
\newcommand{\Phpms}{\hat{P}_{\pm}^S}
\newcommand{\PhXs}{\hat{P}_{X}^S}

\newcommand{\PtdEls}{\wt{P}_{\delta E,L}^S}
\newcommand{\PtgEls}{\wt{P}_{gE,L}^S}
\newcommand{\PtEEls}{\wt{P}_{EE,L}^S}
\newcommand{\PtBBls}{\wt{P}_{BB,L}^S}
\newcommand{\PtXls}{\wt{P}_{X,L}^S}

\newcommand{\PdEls}{P_{\delta E,\ell}^S}
\newcommand{\PgEls}{P_{gE,\ell}^S}
\newcommand{\PEEls}{P_{EE,\ell}^S}

\newcommand{\PgEs}{P_{gE}^S}
\newcommand{\PgBs}{P_{gB}^S}
\newcommand{\PEEs}{P_{EE}^S}
\newcommand{\PBBs}{P_{BB}^S}
\newcommand{\Pggs}{P_{gg}^S}
\newcommand{\Pgps}{P_{g+}^S}
\newcommand{\Pgms}{P_{g\times}^S}
\newcommand{\Ppms}{P_{\pm}^S}
\newcommand{\Pppmms}{P_{+\times}^S}
\newcommand{\Pmmpps}{P_{\times+}^S}
\newcommand{\Pps}{P_{+}^S}
\newcommand{\Pms}{P_{-}^S}

\newcommand{\Ppps}{P_{++}^S}
\newcommand{\Pmms}{P_{\times\times}^S}
\newcommand{\PXs}{P_{X}^S}

\newcommand{\xitXls}{\wt{\xi}_{X,L}^S}
\newcommand{\xitdpls}{\wt{\xi}_{\delta +,L}^S}
\newcommand{\xitgpls}{\wt{\xi}_{g+,L}^S}
\newcommand{\xitmls}{\wt{\xi}_{-,L}^S}

\newcommand{\xiggls}{\xi_{gg,\ell}^S}
\newcommand{\xidpls}{\xi_{\delta +,\ell}^S}
\newcommand{\xigpls}{\xi_{g+,\ell}^S}
\newcommand{\xipls}{\xi_{+,\ell}^S}
\newcommand{\ximls}{\xi_{-,\ell}^S}

\newcommand{\ximpls}{\xi_{\mp,\ell}^S}

\newcommand{\xiXls}{\xi_{X,\ell}^S}

\newcommand{\Pggls}{P_{gg,\ell}^S}

\newcommand{\Ppls}{P_{+,\ell}^S}

\newcommand{\PXls}{P_{X,\ell}^S}

\newcommand{\PXlms}{P_{\ell m}^X}

\newcommand{\Pgplms}{P_{\ell m}^{g+}}

\newcommand{\Pmlms}{P^-_{\ell m}}

\newcommand{\xiXlms}{\Xi_{\ell m}^X}

\newcommand{\Ylm}{Y_{\ell m}}
\newcommand{\Ylma}{Y_{\ell m}^{*}}

\newcommand{\bftheta}{{\boldsymbol \theta}}
\newcommand{\himsun}{{\hbox {$~h^{-1}$}{M_\odot}}}
\newcommand{\himpc}{{\hbox {$~h^{-1}$}{\rm ~Mpc}}}
\newcommand{\higpc}{{\hbox {$~h^{-1}$}{\rm ~Gpc}}}
\newcommand{\hmpci}{{\hbox {$~h{\rm ~Mpc}^{-1}$}}}

\newcommand{\class}{\texttt{\footnotesize CLASS} }

\newcommand{\rockstar}{\texttt{\footnotesize ROCKSTAR} }
\newcommand{\darkquest}{\texttt{D{\footnotesize ARK} Q{\footnotesize UEST} }}

\newcommand{\asiaa}{Academia Sinica Institute of Astronomy and Astrophysics (ASIAA), No. 1, Section 4, Roosevelt Road, Taipei 10617, Taiwan}
\newcommand{\ipmu}{Kavli Institute for the Physics and Mathematics of the Universe (WPI), UTIAS, The University of Tokyo, Kashiwa, Chiba 277-8583, Japan}
\newcommand{\yitp}{Center for Gravitational Physics and Quantum Information, Yukawa Institute for Theoretical Physics, Kyoto University, Kyoto 606-8502, Japan}

\newcommand{\mpa}{Max-Planck-Institut f$\ddot{u}$r Astrophysik, Karl-Schwarzschild-Str. 1, 85741 Garching, Germany}
\newcommand{\kyosan}{Department of Astrophysics and Atmospheric Sciences, Faculty of Science, Kyoto Sangyo University, Kyoto 603-8555, Japan}

\begin{document}

\title
{
Nonlinear redshift space distortion in halo ellipticity correlations: 
\\
Analytical model and $N$-body simulations
}


\author{Teppei Okumura}\email{tokumura@asiaa.sinica.edu.tw}
\affiliation{\asiaa}
\affiliation{\ipmu}

\author{Atsushi Taruya}
\affiliation{\yitp}
\affiliation{\ipmu}

\author{Toshiki Kurita}
\affiliation{\ipmu}
\affiliation{\mpa}

\author{Takahiro Nishimichi}
\affiliation{\kyosan}
\affiliation{\yitp}
\affiliation{\ipmu}

\date{\today} 

\begin{abstract}
We present an analytic model of nonlinear correlators of galaxy/halo
ellipticities in redshift space.  The three-dimensional ellipticity
field is not affected by the redshift-space distortion (RSD) at linear
order, but by the nonlinear one, known as the Finger-of-God effect,
caused by the coordinate transformation from real to redshift space.
Adopting a simple Gaussian damping function to describe the nonlinear
RSD effect and the nonlinear alignment model for the relation between
the observed ellipticity and underlying tidal fields, we derive
analytic formulas for the multipole moments of the power spectra of
the ellipticity field in redshift space expanded in not only the
associated Legendre basis, a natural basis for the projected galaxy
shape field, but also the standard Legendre basis, conventionally used
in literature.  The multipoles of the corresponding correlation
functions of the galaxy shape field are shown to be expressed by a
simple Hankel transform, as is the case for those of the conventional
galaxy density correlations.  We measure these multipoles of the power
spectra and correlation functions of the halo ellipticity field using
large-volume $N$-body simulations. We then show that the measured
alignment signals can be better predicted by our nonlinear model than
the existing linear alignment model.  The formulas derived here have
already been used to place cosmological constraints using from the
redshift-space correlation functions of the galaxy shape field
measured from the Sloan Digital Sky Survey \cite{Okumura:2023}.
\end{abstract}

\maketitle

\section{Introduction} \label{sec:intro}

Intrinsic alignments (IAs) of orientations of galaxies with the
surrounding large-scale structure are considered a main source of
systematics in cosmological gravitational lensing surveys
\cite{Heavens:2000,Croft:2000,Lee:2000,Pen:2000,Catelan:2001,Crittenden:2002,Jing:2002,Hirata:2004,Heymans:2004,Mandelbaum:2006,Hirata:2007,Okumura:2009,Joachimi:2011,Li:2013,Singh:2015,Tonegawa:2018,Tonegawa:2022}
(see also
\cite{Troxel:2015,Joachimi:2015,Kirk:2015,Kiessling:2015,Lamman:2024}
for reviews).  The IA effect has also been attracting attention as a
cosmological probe complimentary to the conventional galaxy
clustering.
It was pointed out that measurements of IAs in three dimensions can be
used as dynamical and geometric probes of cosmology, with
redshift-space distortions and baryon acoustic oscillations (BAO)
\citep{Chisari:2013,Okumura:2019,Taruya:2020,Okumura:2022,Chuang:2022}. Further
theoretical studies have shown that the measurements can be also used
as probes of primordial non-Gaussianity
\citep{Schmidt:2015,Chisari:2016,Akitsu:2021a}, gravitational waves
\citep{Schmidt:2014,Chisari:2014,Biagetti:2020,Akitsu:2023}, neutrino
masses \citep{Lee:2023}, statistical isotropy \citep{Shiraishi:2023}
and gravitational redshifts \citep{Zwetsloot:2022,Saga:2023}.
Recently, observational constraints on cosmological models have been
placed by measuring IAs of galaxies from the Sloan Digital Sky Survey
(SDSS) \citep{Okumura:2023,Kurita:2023,Xu:2023}.  More significant
contributions on the cosmological constraints are expected by
observations of the IA of galaxies in ongoing and upcoming galaxy
redshift surveys with a better imaging quality
\cite{Takada:2014,DESI-Collaboration:2016,Euclid_Collaboration:2020}.

In order to maximize the cosmological information encoded in the IA of
galaxies, one needs to develop accurate non-linear models of IA
statistics in full three dimensions.  While modeling of the nonlinear
power spectrum in redshift has been extensively performed for the
galaxy density field
\cite[e.g.,][]{Peacock:1994,Park:1994,Scoccimarro:2004,Matsubara:2008,Taruya:2010,Seljak:2011,Okumura:2015},
there are fewer studies for the galaxy ellipticity field.  The
simplest model for the IA statistics is the linear alignment (LA)
model, which linearly relates the ellipticity field to the tidal
gravitational field
\cite{Catelan:2001,Hirata:2004,Okumura:2019,Okumura:2020}.  The model
beyond the LA model, nonlinear alignment (NLA) model as well as the
nonlinear shape bias model have been studied
\cite{Blazek:2015,Schmitz:2018,Blazek:2019,Vlah:2020,Taruya:2021,Matsubara:2022,Matsubara:2022a,Matsubara:2023,Bakx:2023,Akitsu:2023a,Maion:2023}.
However, the modeling of the IA statistics in three dimensions
requires an understanding of the nonlinear redshift-space distortion
(RSD) effect on them \cite{Jackson:1972,Kaiser:1987}, which is not as
trivial as the nonlinearities above:
Refs.~\cite{Blazek:2011,Chisari:2013} had included the linear Kaiser
factor into the redshift-space ellipticity field but later
Ref.~\cite{Singh:2015} pointed out that the ellipticity field is not
affected by the linear RSD.  This argument is no longer valid once
nonlinearities of IA are taken into account.  It has been shown by
Refs.~\cite{Matsubara:2022a,Chen:2024} in their analytical models that
the ellipticity field is indeed affected by the nonlinear RSD
effect. The accuracy of the models including the nonlinear RSD effect
is still unclear because they have not been tested against the
measurements from $N$-body simulations (but see
Ref.~\cite{Okumura:2023} which provided the direct comparison with the
observed result).

In this paper, we derive formulas of nonlinear intrinsic alignment
statistics in redshift space where the nonlinear RSD effect is taken
into account as a Gaussian damping function. We provide
phenomenological formulas of both the IA power spectra and correlation
functions.  Conventionally, the IA correlations are expanded in terms
of the (standard) Legendre polynomials
\cite{Singh:2016,Okumura:2020,Kurita:2021}.  However, it was pointed
out by Ref.~\cite{Kurita:2022} that the IA statistics which contain
the the geometric factor due to the projection of the intrinsic shape
along the line of sight are more naturally expanded in terms of the
associated Legendre polynomials rather than the standard Legendre ones
(see also Ref.~\citep{Saga:2023,Shi:2024,Singh:2023}).  Thus, we
provide our formulas expanded by both standard and associated Legendre
polynomials.  Note that the formulas derived here have already been
used to place cosmological constraints from the Sloan Digital Sky
Survey in Ref.~\cite{Okumura:2023}.

Before proceeding to the next section, we note that we assume the
linear and scale-independent bias for both the density and shape
fields throughout the paper.  The nonlinear shape bias has been
investigated in real space
(e.g.,~\cite{Schmitz:2018,Akitsu:2021,Bakx:2023,Akitsu:2023a}).
Incorporating the nonlinear bias effect into theoretical predictions
of redshift-space IA statistics will be presented in the future work
(but see Ref.~\cite{Taruya:2021} for the impacts of nonlinear bias on
the super-sample effects).

The rest of this paper is organized as follows.  In
Section~\ref{sec:preliminaries} we describe general forms of the
nonlinear power spectra of the ellipticity field in redshift space.
Sections~\ref{sec:nl_model} and \ref{sec:nl_2pcf} present the
analytical model for the multipole moments of the IA power spectra and
correlation functions, taking into account the nonlinear RSD and
alignment effects.  In Section~\ref{sec:analytic_result} we presents
the results of numerical calculations for the nonlinear RSD model of
IA statistics.  Our models for the IA statistics are tested against
the measurements from $N$-body simulations in Section
\ref{sec:result}.  Our conclusions are given in
Section~\ref{sec:conclusion}.  In Appendix \ref{sec:derivation}, we
provide alternative ways to derive the model of the IA correlation
functions by using the spherical harmonic expansion.  The higher-order
terms that do not contain linear contributions are provided in
Appendix \ref{sec:higher_order}.  The expressions at the linear theory
limit are provided in Appendix \ref{sec:linear_limit}.

Throughout the paper, we assume the spatially flat $\Lambda$CDM model
as our fiducial model \cite{Planck-Collaboration:2016}: $\Omegam =
1-\OmegaDE = 0.315$, $\OmegaK=0$, $w_0 = -1$, $w_a=0$, $H_0=67.3~[{\rm
    km/s/Mpc}]$ and the present-day value of $\sigma_8$ to be
$\sigma_8 = 0.8309$.

\section{Galaxy density and ellipticity power spectra in redshift space}\label{sec:preliminaries}

In this paper, we consider redshift-space correlators of the
galaxy/halo ellipticity field with itself as well as with the density
field.  The positions of objects in three-dimensional galaxy surveys
are sampled by their redshifts and are therefore displaced along the
line of sight by their peculiar velocities, known as redshift-space
distortions (RSD).
The position of distant objects in real space, $\bfx$, is mapped to
the one in redshift space, $\bfs$, as $\bfs = \bfx +
\frac{v_z(\bfx)}{aH(z)}\hat{\bfz}$, where $v_z(\bfx)=\bfv(\bfx)\cdot
\hat{\bfz}$, $\bfv(\bfx)$ is the peculiar velocity, $H(z)$ is the
Hubble parameter at redshift $z$, $a=(1+z)^{-1}$ is the scale factor,
and hat denotes a unit vector and $\hat{\bfz}$ is pointing the
observer's line of sight, namely,
$\hat{\bfz}=\hat{\bfs}=\hat{\bfx}$.\footnote{ While properly taking
into account the wide-angle effect \cite{Szalay:1998,Szapudi:2004}
provides additional cosmological constraints (see,
Refs. \cite{Shiraishi:2021a,Saga:2023} for the studies of the
wide-angle effects on ellipticity fields), throughout this paper, we
assume a plane-parallel approximation because we mainly focus on the
correlations at small separations.  }

\subsection{Galaxy density field \label{sec:galaxy_density}}

Density perturbations of galaxies/halos are defined by 
the density contract from the mean $\bar{\rho}_g$, 
\be
\delta_g(\bfx) \equiv \rho_g(\bfx) / \bar{\rho}_g - 1. \label{eq:delta}
\ee
Through the real-to-redshift space mapping, the redshift-space density
field of galaxies is given by
\citep{Scoccimarro:2004,Taruya:2010,Seljak:2011}:
\begin{align}
1+\deltags(\bfs) = \int d^3\bfx \int \frac{d^3\bfk}{(2\pi)^3} e^{i \bfk \cdot [\bfs-\bfx+f u_z(\bfx) \hat{\bfz}]}
\nn \\ \times 
[ 1+\deltag(\bfx)],
\end{align}
where $u_z (\bfx)= -v_z(\bfx)/(faH)$ and the superscript $S$ denotes a
quantity defined in redshift space.
The quantity $f$ is the growth rate parameter to characterize the
evolution of the density perturbation, defined as
\be
f(z) = -\frac{d\ln D(z)}{d \ln (1+z)} =  \frac{d\ln D(a)}{d \ln a}, \label{eq:fz}
\ee
where $D(z)$ is the linear growth factor of the perturbation,
$D(z)=\delta_m(\bfx;z) / \delta_m(\bfx;0)$.
Via the Fourier transform,
\begin{align}
\deltags(\bfk) & = \int d^3\bfs e^{-i\bfk\cdot\bfs}\deltags(\bfs) \nn \\
& =\int d^3\bfx e^{-i\bfk\cdot[\bfx-f u_z(\bfx)\hat{\bfz}]}\left[\deltag(\bfx)+f \nabla_z u_z(\bfx)\right]. \label{eq:delta_k}
\end{align}
The galaxy auto-power spectrum, $\Pggs(\bfk)$, with the most general
form under the plane-parallel approximation can be written as
\cite{Scoccimarro:1999,Scoccimarro:2004}
\begin{align}
\Pggs(\bfk)& =\int d^3\bfr e^{-i\bfk\cdot\bfr}\bigl\langle e^{ifk\mu_{\bfk}\Delta u_z} \bigr. \nn \\
&\bigl. \times \left[\delta_g(\bfx)+f\nabla_z u_z(\bfx)\right] \left[\delta_g(\bfx')+f\nabla_z u_z(\bfx')\right]\bigr\rangle, \label{eq:Pgg_general}
\end{align}
where $\bfr = \bfx - \bfx'$, $\Delta u_z = u_z(\bfx)-u_z(\bfx')$ and
$\mu_{\bfk}=\hat{\bfk}\cdot\hat{\bfx}$ is the directional cosine
between the wave vector and line-of-sight direction.

\subsection{Galaxy intrinsic ellipticity field \label{sec:galaxy_ellipticity}}

We use ellipticities of galaxies as a tracer of the 3-dimensional
tidal field.  Similarly to the density field, the redshift-space
ellipticity field is described as (See Ref.~\cite{Okumura:2014} for a
similar equation for the redshift-space velocity field)
\begin{align}
\gammaijs(\bfs) = \int d^3\bfx \int \frac{d^3\bfk}{(2\pi)^3} &e^{i \bfk \cdot [\bfs-\bfx+fu_z (\bfx)\hat{\bfz}]} 
\gamma_{ij}(\bfx),\label{eq:gamma_s}
\end{align}
which is translated into Fourier space as
\begin{align}
\gammaijs(\bfk)  &= \int d^3\bfs e^{-i\bfk\cdot\bfs}\gammaijs(\bfs)
 \nn \\ & 
=\int d^3\bfx e^{-i\bfk\cdot[\bfx-fu_z(\bfx)\hat{\bfz}]}\gamma_{ij}(\bfx). \label{eq:gamma_k}
\end{align}
The auto correlation of the ellipticity field and its cross
correlation with the density field are, respectively,
\begin{align}
P_{\gamma\gamma,ijkl}^S(\bfk) = & \int d^3\bfr e^{-i\bfk\cdot\bfr}\bigl\langle e^{ifk\mu_{\bfk}\Delta u_z} \gamma_{ij}(\bfx)\gamma_{kl}(\bfx')\bigr\rangle, \label{eq:Pgammagamma_general} \\
P_{g\gamma,ij}^S(\bfk) = & \int d^3\bfr e^{-i\bfk\cdot\bfr}\bigl\langle e^{ifk\mu_{\bfk}\Delta u_z} \bigr. \nn \\
&\qquad \quad \bigl. \times \left[\delta_g(\bfx)+f\nabla_z u_z(\bfx)\right] \gamma_{ij}(\bfx') \bigr\rangle. \label{eq:Pggamma_general}
\end{align}
Note that $\gammaijs$ in Eq.~(\ref{eq:gamma_s}) can be interpreted as
either the volume-weighted or number-weighted ellipticity
field. Accordingly, Eqs~(\ref{eq:Pgammagamma_general}) and
(\ref{eq:Pggamma_general}) are the power spectra for the
volume-weighted or number-weighted ellipticity field.

Since the observed galaxy/halo shapes are projected onto the celestial sphere ($x-y$ plane), 
we consider the two traceless components as the observed shape field,
\begin{align}
\left(
\begin{array}{c} \gamma_+^S(\bfk) \\ \gamma_\times^S(\bfk) \end{array}
\right) =
\left(
\begin{array}{c} 
    \gamma_{xx}^S(\bfk)-\gamma_{yy}^S(\bfk) \\
    2\gamma_{xy}^S(\bfk)
\end{array}
\right).
\end{align}
All the power spectra with this projected shape field can be expressed
in terms of $P_{\gamma\gamma,ijkl}^S$
(Eq.~\ref{eq:Pgammagamma_general}) and $P_{g\gamma,ij}^S$
(Eq.~\ref{eq:Pggamma_general}), as
\begin{align}
\Pgps(\bfk) &= P_{g\gamma,xx}^S(\bfk) - P_{g\gamma,yy}^S(\bfk),  \label{eq:Pgp_general} \\
\Pgms(\bfk) &= 2P_{g\gamma,xy}^S(\bfk), \\
\Ppps(\bfk)& = P_{\gamma\gamma,xxxx}^S(\bfk) - 2P_{\gamma\gamma,xxyy}^S(\bfk) \nn \\
&\qquad \qquad + P_{\gamma\gamma,yyyy}^S(\bfk),\\
\Pmms(\bfk)&=4P_{\gamma\gamma,xyxy}^S(\bfk), \\
\Pppmms(\bfk)&=\Pmmpps(\bfk) \nn \\
&= 2\left[ P_{\gamma\gamma,xxxy}^S(\bfk)-P_{\gamma\gamma,yyxy}^S(\bfk) \right],\label{eq:Pppmm_general}\\
\Ppms(\bfk)&=\Ppps(\bfk)\pm\Pmms(\bfk), \label{eq:Ppm_general}
\end{align}

We also define E-/B-modes, $\gamma_{(E,B)}$, which are the
rotation-invariant decomposition of the ellipticity field
\cite{Crittenden:2002},
\begin{align}
\gamma_E^S(\bfk) + i\gamma_B^S(\bfk) = e^{-2i\phi_k}\left[  \gamma_{+}^S(\bfk)+i\gamma_{\times}^S(\bfk) \right],
\label{eq:gamma_E_deltam}
\end{align}
where $\phi_k$ is the azimuthal angle of the wavevector projected on
the celestial sphere.
Then the power spectra of the E-/B-modes are expressed using those of $\gamma_{(+,\times)}^S(\bfk)$  [Eqs.~(\ref{eq:Pgp_general}) -- (\ref{eq:Pppmm_general})]:
\begin{align}
\PgEs(\bfk)& = \cos{(2\phi_{\bfk})}\Pgps(\bfk) + \sin{(2\phi_{\bfk})}\Pgms(\bfk) ,\label{eq:PgE_general} \\
\PgBs(\bfk) &= -\sin{(2\phi_{\bfk})}\Pgps(\bfk) + \cos{(2\phi_{\bfk})}\Pgms(\bfk) ,\\
\PEEs(\bfk)&=\cos^2{(2\phi_{\bfk})}\Ppps(\bfk)+ \sin^2{(2\phi_{\bfk})}\Pmms(\bfk) \nn \\ 
& \qquad +2\cos{(2\phi_{\bfk})}\sin{(2\phi_{\bfk})}\Pppmms(\bfk), \\
\PBBs(\bfk)&=\sin^2{(2\phi_{\bfk})}\Ppps(\bfk)+ \cos^2{(2\phi_{\bfk})}\Pmms(\bfk) \nn \\ 
& \qquad -2\cos{(2\phi_{\bfk})}\sin{(2\phi_{\bfk})}\Pppmms(\bfk). \label{eq:PBB_general}
\end{align} 

Comparison of Eqs.~(\ref{eq:Pgammagamma_general}) and
(\ref{eq:Pggamma_general}) with Eq.~(\ref{eq:Pgg_general}) implies
that the power spectra of the shape field in redshift space is
affected by the Finger-of-God effect in the same way as those of the
density field. We investigate the effect in the next section.

\section{Analytical model of nonlinear redshift-space distortion}\label{sec:nl_model}

\subsection{Nonlinear alignment model}\label{sec:la_model}

To relate the observed galaxy/halo shape field to the underlying tidal
gravitational field, we use the linear alignment (LA) model, which
assumes a linear relation between the intrinsic ellipticity and tidal
field at the true three-dimensional position, namely without the
line-of-sight displacement due to RSDs
\citep{Catelan:2001,Hirata:2004}.  In Fourier space, the ellipticity
field projected along the line of sight, $\hat{\bfz}$, is given by
\begin{align}
\left(
\begin{array}{c} \gamma_+(\bfk) \\ \gamma_\times(\bfk) \end{array}
\right) =b_K
\left(
\begin{array}{c} 
    k_{x}^{2}-k_{y}^{2} \\
    2k_{x}k_{y}
\end{array}
\right)
\frac{\delta_m(\bfk)}{k^2}, \label{eq:shape_proj}
\end{align}
where $b_{K}$ represents the redshift-dependent coefficient of the
intrinsic alignments which we refer to as the shape bias\footnote{The
shape bias parameter is related to $\wt{C}_1$ in
Ref.~\cite{Okumura:2020} by $b_K = -\wt{C}_1$. Some references use
$b_K$ for the shape bias in the three-dimensional tidal field,
$K_{ij}$, as $\gamma_{ij}=b_K K_{ij}$ (e.g.,
\cite{Schmidt:2015}). However, this $b_K$ differs from that we
introduced in Eq.~(\ref{eq:shape_proj}) by a factor of two.}.
We adopt the nonlinear alignment (NLA) model, which replaces the
linear matter density field $\delta_m$ by the nonlinear one
\citep{Bridle:2007}. Furthermore, the redshift-space ellipticity field
is multiplied by the damping function due to the nonlinear RSD effect
as we will see below.


\subsection{Phenomenological RSD model}\label{sec:fog}
To accurately describe the observed and simulated results of
galaxy/halo ellipticity correlations in redshift space, the nonlinear
RSD effect, known as the Finger-of-God (FoG), needs to be taken into
account in the theoretical predictions. In
Sec.~\ref{sec:preliminaries}, we see that the expressions of power
spectra involve the exponential factor in the ensemble average [see
  Eqs.~\eqref{eq:Pgg_general}, \eqref{eq:Pgammagamma_general} and
  \eqref{eq:Pggamma_general}], which is responsible for suppressing
the power spectrum amplitude due to the randomness of the pair-wise
velocity contributions. Although the impact of this exponential
factor, coupled with the density and ellipticity fields, needs to be
carefully treated in the statistical calculations, a dominant effect
of this is phenomenologically but quantitatively described by imposing
the factorized ansatz, i.e., $\langle e^{-ifk\mu_k\Delta
  u_z}\{\cdots\}\rangle \to \langle e^{-ifk\mu_k\Delta u_z}\rangle
\langle\cdots\rangle$. Then, writing all the redshift-space power
spectra in the previous subsection as $\PXs(\bfk)$, this ansatz leads
to the following separable expression of the power spectra:
\cite{Scoccimarro:2004} (see also
\cite{Peacock:1994,Park:1994,Taruya:2009,Okumura:2012}),
\begin{align}
\PXs(\bfk) = \PhXs(\bfk)\Dfog^2(fk\mu_{\bfk}\sigv), \label{eq:redshift_power}
\end{align}
where $k=|\bfk|$, $\mu_{\bfk}= \hat\bfk \cdot \hat\bfz=k_z / k$ is the
directional cosine between the observer's line of sight and the
wavevector $\bfk$. In this expression, the exponential factor $\langle
e^{-ifk\mu_{\bfk}\Delta u_z}\rangle$ is identified with the function
$D_{\rm FoG}^2$, and we model it as the zero-lag correlation
characterized by the velocity dispersion, $\sigma_{\rm v}$.

Eq.~\eqref{eq:redshift_power} includes the redshift-space galaxy power
spectrum proposed by Ref.~\cite{Scoccimarro:2004} ($X=gg$). In this
case, the function $\Phggs$ is given by
\begin{align}
& \Phggs (\bfk) = 
b^2 \Pdd (k)   + 2bf\mu_{\bfk}^2 \Pdt (k)  + f^2\mu_{\bfk}^4 \Ptt(k) 
,  \label{eq:scoccimarro} 
\end{align}
where $b$ is the linear galaxy bias \citep{Kaiser:1984}, $\Pdd$ and
$\Ptt$ are the nonlinear auto-power spectra of density and velocity
divergence, respectively, and $\Pdt$ is the their cross-power
spectrum. In the linear theory limit, $\Pdd = \Pdt = \Ptt$ and
$\Dfog=1$, and hence Equation~(\ref{eq:scoccimarro}) converges to the
original Kaiser formula \cite{Kaiser:1987}.  The parameter $f$
quantifies the cosmological velocity field and the speed of structure
growth, and thus is useful for testing a possible deviation of the
gravity law from general relativity \citep{Guzzo:2008,Okumura:2016}.
Under modified gravity models, even though the background evolution is
the same as the $\Lambda$CDM model, the density perturbations would
evolve differently (See, e.g., Ref. \cite{Matsumoto:2020} for
degeneracies between the expansion and growth rates for various
gravity models).  The relation of this phenomenological form with the
general one in Eq.~(\ref{eq:Pgg_general}) was made in
Ref.~\cite{Taruya:2010}.

Adopting the NLA model, the cross-power spectra of the galaxy density
and E-/B-mode fields and the auto-power spectra of the E-/B-mode
fields are given by
\begin{align}
& \PhgEs (\bfk)  = b_K (1-\mu_{\bfk}^2)\left[b\Pdd(k)+f\mu_{\bfk}^2\Pdt(k)\right]  ,
\label{eq:PhgE}\\
& \PhEEs (\bfk)  = b_K^2 (1-\mu_{\bfk}^2)^2 \Pdd(k) ,
\label{eq:PhEE}\\
&\PhgBs(\bfk) = \PhBBs(\bfk)=0.
\end{align}
The power spectra of the E-mode, namely $\PgEs$ and $\PEEs$, are the
main statistics to be tested in this paper.  Our model for these
statistics contain free parameters of $\bftheta=(f,b,b_K, \sigv)$.

When we analyze power spectra that are anisotropic along the line of
sight and thus have $\mu_{\bfk}$ dependences, we commonly use the
multipole expansion in terms of the Legendre polynomials, ${\cal
  L}_\ell(\mu_{\bfk})$, as \cite{Hamilton:1992,Okumura:2020}:
\begin{align}
&\PXs(\bfk)= \sum_{\ell}\PXls(k){\cal L}_{\ell}(\mu_{\bfk}), \label{eq:standard_legendre}
\end{align}
where the coefficients $\PXls(k)$ are given by
\begin{align}
&\PXls(k) = \frac{2\ell+1}{2}\int^{1}_{-1} d\mu_{\bfk} \PXs(k,\mu_{\bfk}) \mathcal{L}_\ell (\mu_{\bfk}).
\end{align}

In this paper, we adopt a simple Gaussian function for the nonlinear
RSD term,
$\Dfog(fk\mu_{\bfk}\sigv) =\exp{\left[-(fk\mu_{\bfk}\sigv)^2/2\right]}
\equiv \exp{\left(-\alpha \mu_{\bfk}^2/2\right)}$, where
$\alpha=(fk\sigv)^2$ .
With this Gaussian function, all the multipole moments of the power
spectra, i.e., $\PXls(k)$, are expressed in a factorized form with the
angular dependence encoded by
\begin{align}
p^{(n)}(\alpha)\equiv \int^1_{-1} d\mu_{\bfk} \mu_{\bfk}^{2n} e^{-\alpha\mu_{\bfk}^2}
. \label{eq:pnalpha}
\end{align}
This function has a maximum at $\alpha=0$ and decreases monotonically.
The integral can be performed analytically as $p^{(n)}(\alpha)=
\frac{\gamma(1/2+n,\alpha)}{\alpha^{1/2+n}},$ where $\gamma(n,\alpha)
= \int^\alpha_0 dt \ t^{n-1}e^{-t}$ is the incomplete gamma function
of the first kind.  All the formulas derived below are, thus,
expressed in terms of the function $p^{(n)}(\alpha)$. The
linear-theory limits of the formulas are derived by setting
$\alpha=0$, $p^{(n)}(0)=\frac{2}{2n+1}$.

The expression of the multipole expansions for the nonlinear galaxy
power spectra with the Gaussian damping function were derived by
Refs.~\cite{Taruya:2009,Percival:2009}.  They are expressed in terms
of $p^{(n)}(\alpha)$ as
\begin{align}
\Pggls(k) &= b^2\mathcal{Q}_{gg,\ell}^{(0)}(\alpha)\Pdd(k) + 2bf \mathcal{Q}_{gg,\ell}^{(1)}(\alpha)\Pdt(k) \nn \\ 
& \qquad\qquad\qquad + f^2\mathcal{Q}_{gg,\ell}^{(2)}(\alpha)\Ptt(k). \label{eq:Pggls}
\end{align}
The multipoles with $\ell=0,2,4$ contain non-zero contributions at
linear scales.  The functions $\mathcal{Q}_{gg,\ell}^{(n)}(\alpha)$ up
to $\ell=4$ are given by
\begin{align}
\mathcal{Q}_{gg,0}^{(n)}(\alpha) &= \frac12 p^{(n)}(\alpha),\\
\mathcal{Q}_{gg,2}^{(n)}(\alpha) & = \frac54 \left[3 p^{(n+1)}(\alpha) - p^{(n)}(\alpha)\right], \\
\mathcal{Q}_{gg,4}^{(n)}(\alpha) & = \frac{9}{16} \left[35 p^{(n+2)}(\alpha) - 30p^{(n+1)}(\alpha) 
\right. \nn \\ & \qquad\qquad\qquad\qquad\quad \left. 
+3p^{(n)}(\alpha) \right].
\end{align}
By taking the $\alpha \to 0$ limit, $\mathcal{Q}_{gg,\ell}(0) = 0$ for
$\ell>4$ and the linear RSD formulas are obtained \cite{Kaiser:1987}.


\subsection{E-mode cross- and auto-power spectra}

Due to the geometric factor, $(1-\mu_{\bfk}^2)$, arising from the
projection of the shape field along the line of sight in
Eqs.~(\ref{eq:PhgE}) and (\ref{eq:PhEE}), the cross-power spectra of
the galaxy density and E-mode shape fields (gE power) and auto-power
spectra of the E-mode shape field (EE power) are more naturally
expressed in the associated Legendre basis rather than in the standard
Legendre basis,
\begin{align}
&\PXs(\bfk)= \sum_{L\geq m}\PtXls(k)\Theta_{L}^{m}(\mu_{\bfk}), \label{eq:PXs_normalized}
\end{align}
where $X=\{gE, EE\}$, 
$\Theta_{L}^{m}(\mu)$ is the normalized associated Legendre function related to the unnormalized one by
$\Theta_{L}^{m}(\mu) =\sqrt{\frac{2L+1}{2}\frac{(L-m)!}{(L+m)!}}\mathcal{L}_{L}^{m}(\mu_{\bfk})$,
so that it satisfies a simple orthonormal relation,
$
\int^1_{-1}d\mu\Theta_{L}^{m}(\mu)\Theta_{L'}^{m}(\mu)=\delta_{LL'},
$
with $\delta_{LL'}$ being the Kronecker's delta.  We added tilde to
$P_{X,L}^S(k)$ to emphasize that they are the expansion coefficients
of the normalized polynomials $\Theta_{L}^{m}(\mu_{\bfk})$.  While the
choice of $m$ in Eq.~(\ref{eq:PXs_normalized}) is arbitrary, the
expressions of the gE and EE power spectra become the simplest if one
chooses $m=2$ and $4$, respectively \cite{Kurita:2022}.  Thus, in the
following, $\PtgEls(k)$ and $\PtEEls(k)$ stand for the coefficients
expanded by $\Theta_L^{m=2}$ and $\Theta_L^{m=4}$, respectively.
We also provide the expression of the multipoles of the gE and EE
power spectra expanded in terms of the standard Legendre polynomials,
which used to be commonly considered theoretically but were found to
be not direct observables in real surveys \cite{Kurita:2022}.

All the multipoles of the power spectra considered in this subsection
are summarized in Table \ref{tab:pk}.  We present only the multipoles
$\PtXls$ and $\PXls$ that contain linear-order contributions here, and
the higher-order terms are provided in Appendix \ref{sec:higher_order}
up to $L=12$ and $\ell = 12$, respectively.  We can obtain the linear
theory expressions by taking the $\alpha\to 0$ limit, and they are
shown in Appendix \ref{sec:linear_limit}.


\begin{table}[tb]
\caption{Summary of the coordinate-independent alignment power spectra
  derived in section \ref{sec:nl_model}.  The functions $\PtXls$ and
  $\PXls$ are coefficients expanded in terms of the normalized
  associated Legendre polynomial $\Theta_L$ and standard Legendre
  polynomial $\mathcal{L}_\ell$, respectively. }
\begin{center}
\begin{tabular}{lcccccccc}
\hline\hline
Statistics  & & Definition & & \multicolumn{2}{l}{Fourier-space} & &\multicolumn{2}{c}{Result (Fig.)} \vspace{-1mm} \\ 
                                                &&      (Eq.)       & &     multipole        &  (Eq.) && Theory & Sim.  \\ 
\noalign{\hrule height 1pt}
gE &&$\PgEs(\bfk) $                  (\ref{eq:PhgE}) &&$\PtgEls$ & (\ref{eq:PtgEls}) && \ref{fig:pkxi_lm_theory} & \ref{fig:pk_lm_red}  \\
                             &&                                          &&$\PgEls$ & (\ref{eq:PgEls}) && \ref{fig:pkxi_l_theory} & \ref{fig:pk_l_red}   \\
EE &&$\PEEs(\bfk)$                  (\ref{eq:PhEE}) && $\PtEEls$ & (\ref{eq:PtEEls}) &&\ref{fig:pkxi_lm_theory} & \ref{fig:pk_lm_red} \\ 
                             &&                                          &&$\PEEls$  & (\ref{eq:PEEls}) && \ref{fig:pkxi_l_theory}& \ref{fig:pk_l_red} \\
\hline\hline
\end{tabular}
\end{center}
\label{tab:pk}
\end{table}

Using the function $p^{(n)}(\alpha)$ [Eq.~(\ref{eq:pnalpha})], the gE
power spectrum, $\PtgEls(k)$, is explicitly described as
\begin{align}
\PtgEls(k)&=\int^1_{-1}d\mu_{\bfk}\PgEs(k,\mu_{\bfk})\Theta_{L}^{m=2}(\mu_{\bfk}) \nn \\
&= bb_K\mathcal{\wt{Q}}_{gE,L}^{(0)}(\alpha)\Pdd(k) 
\nn \\ & \qquad\qquad 
+ fb_K \mathcal{\wt{Q}}_{gE,L}^{(1)}(\alpha)\Pdt(k), \label{eq:PtgEls}
\end{align}
where $L\geq m=2$.  The two lowest multipoles, $\PtgEls(k)$ with $L=2$
and $4$, contain linear-order contributions, and
$\mathcal{Q}_{gE,L}^{(n)}(\alpha)$ are given by
\begin{align}
\mathcal{\wt{Q}}_{gE,2}^{(n)}(\alpha) &=\frac{\sqrt{15}}{4}  \left[ p^{(n)}(\alpha)-2p^{(n+1)}(\alpha) 
\right. \nn \\ & \qquad \qquad \left.
+p^{(n+2)}(\alpha)\right],\\
\mathcal{\wt{Q}}_{gE,4}^{(n)}(\alpha) &=\frac{3\sqrt{5}}{8}  \left[ -p^{(n)}(\alpha) +9p^{(n+1)}(\alpha)
\right. \nn \\ & \qquad \qquad \left.
 -15p^{(n+2)}(\alpha) +7p^{(n+3)}(\alpha) \right].
 \end{align}

When the gE power spectrum is expanded by the standard Legendre basis
[Eq.~(\ref{eq:standard_legendre})] instead of the associated Legendre
basis, they are shown to be:
\begin{align}
\PgEls(k) &=  bb_K\mathcal{Q}_{gE,\ell}^{(0)}(\alpha)\Pdd(k) 
\nn \\ & \qquad\qquad 
+ fb_K \mathcal{Q}_{gE,\ell}^{(1)}(\alpha)\Pdt(k), \label{eq:PgEls}
\end{align}
where $\mathcal{Q}_{gE,\ell}^{(n)}$ which contain the linear contributions are given by
\begin{align}
\mathcal{Q}_{gE,0}^{(n)}(\alpha) &=\frac{1}{2} \left[p^{(n)}(\alpha)-p^{(n+1)}(\alpha)\right],\\
\mathcal{Q}_{gE,2}^{(n)}(\alpha) &=-\frac{5}{4} \left[p^{(n)}(\alpha) -4 p^{(n+1)}(\alpha) 
\right. \nn \\ & \qquad \left.
+ 3 p^{(n+2)}(\alpha) \right],\\
\mathcal{Q}_{gE,4}^{(n)}(\alpha) &=
\frac{9}{16} \left[3p^{(n)}(\alpha)-33p^{(n+1)}(\alpha)
\right. \nn \\ & \qquad \left.
+65p^{(n+2)}(\alpha)-35p^{(n+3)}(\alpha)\right].
\end{align}

Similarly to the gE power spectrum in Eq.~(\ref{eq:PtgEls}), the EE
auto-power spectrum expanded in terms of the associated Legendre
polynomials, $\PtEEls(k)$, is concisely described as
\begin{align}
\PtEEls(k)&=\int^1_{-1}d\mu_{\bfk}\PEEs(k,\mu_{\bfk})\Theta_{L}^{m=4}(\mu_{\bfk}) \nn \\
&= b_K^2 \mathcal{\wt{Q}}_{EE,L} (\alpha)\Pdd(k), \label{eq:PtEEls}
\end{align}
where $L\geq m=4$.  Only the lowest-order coefficient with $L=4$,
$\mathcal{\wt{Q}}_{EE,4}$, contains the contribution in linear theory.
The term is given by
\begin{align}
\mathcal{\wt{Q}}_{EE,4}(\alpha)& =\frac{3 \sqrt{35}}{16}\left[ p^{(0)}(\alpha) -4p^{(1)}(\alpha) +6p^{(2)}(\alpha)
\right. \nn\\& \qquad \qquad  \left. 
-4p^{(3)}(\alpha)+p^{(4)}(\alpha)\right].
\end{align}

The multipole expansions of the EE power in terms of the standard
Legendre polynomials are expressed as
\begin{align}
\PEEls(k) 
&= b_K^2 \mathcal{Q}_{EE,\ell} (\alpha)\Pdd(k), \label{eq:PEEls}
\end{align}
where 
\begin{align}
\mathcal{Q}_{EE,0}(k) &= \frac12\left[p^{(0)}(\alpha) - 2p^{(1)}(\alpha) + p^{(2)}(\alpha)\right],  \label{eq:QEE0}\\
\mathcal{Q}_{EE,2}(k) &= -\frac54 \left[p^{(0)}(\alpha) - 5p^{(1)}(\alpha) + 7 p^{(2)}(\alpha) 
\right. \nn \\ &\qquad \quad \quad \quad  \left.
-3 p^{(3)}(\alpha) \right], \label{eq:QEE2}\\
\mathcal{Q}_{EE,4}(\alpha) &= \frac{9}{16} \left[3p^{(0)}(\alpha) - 36p^{(1)}(\alpha) + 98 p^{(2)}(\alpha)
\right. \nn \\ &\qquad \quad \quad \quad  \left.
-100 p^{(3)} (\alpha) + 35p^{(4)} (\alpha)\right]. \label{eq:QEE4}
\end{align}

Note that, which basis is preferred in the actual cosmological
analysis of the alignment statistics depends on which aspects one
considers more important. As presented above, we can express the IA
power spectra using the minimum set of the multipoles in the
associated Legendre basis. However, cosmological information encoded
at linear level in the standard Legendre basis is propagated into the
higher-order multipoles that have no linear level contribution in the
associated one to some extent due to the off-diagonal components of
the covariance matrix for the power spectra. It is demonstrated in our
upcoming paper \cite{Inoue:2024}.

Numerical results of these nonlinear formulas are given in
Sec.~\ref{sec:analytic_result} and compared to the measurements from
$N$-body simulations in Sec.~\ref{sec:result}.


\begin{table}[tb]
\caption{Same as Table \ref{tab:pk} but for the coordinate-dependent alignment power spectra derived in section \ref{sec:nl_2pcf}.
The power spectra are used to derive the expressions for the corresponding correlation functions, $\xiXls$ and $\xitXls$.}
\begin{center}
\begin{tabular}{lccccccccc}
\hline\hline
Statistics  & & Definition &&  \multicolumn{2}{l}{Configuration-space} &&\multicolumn{2}{c}{Result (Fig.)} \vspace{-1mm} \\ 
   &&      (Eq.)        &&     multipole        &  (Eq.)  && Theory & Sim.  \\ 
\noalign{\hrule height 1pt}
II($+$) &&$\Pps(\bfk) $                  (\ref{eq:pk_iip}) && $\xipls$& (\ref{eq:xipls}) &&& \ref{fig:xiia_l_red}  \\
GI    &&$\Pgps(\bfk)$                (\ref{eq:pk_gi}) && $\xitgpls$ &(\ref{eq:xigp_multipole_associated_legendre}) &&\ref{fig:pkxi_lm_theory} & \ref{fig:xi_lm_red} \\ 
            &&                                                              & &                                   $\xigpls$ &(\ref{xigp0s})--(\ref{xigp4s})&& \ref{fig:pkxi_l_theory}& \ref{fig:xiia_l_red} \\
II($-$)  &&$\Pms(\bfk)$                (\ref{eq:pk_iip}) &&$\xitmls$ &(\ref{eq:xim_multipole_associated_legendre}) &&\ref{fig:pkxi_lm_theory} & \ref{fig:xi_lm_red} \\ 
             &&                                                              & &                                         $\ximls$ &(\ref{xim0s})--(\ref{xim4s})&& \ref{fig:pkxi_l_theory}& \ref{fig:xiia_l_red} \\
\hline\hline
\end{tabular}
\end{center}
\label{tab:xi}
\end{table}

\section{Ellipticity correlation functions}\label{sec:nl_2pcf}

In this section, we present multipole moments of the correlation
functions of the projected galaxy/halo ellipticity field in redshift
space.  First, the model for the II($+$) correlation multipoles in
terms of the standard Legendre polynomials, $\xipls$, is given.
Second, we provide the models for the GI and II($-$) multipoles in
terms of the associated Legendre polynomials, $\xitgpls$ and
$\xitmls$, respectively, and then those of the standard Legendre
polynomials, $\xigpls$ and $\ximls$.  Table \ref{tab:xi} summarizes
all the power spectra and correlation functions of the shape field
considered in this section.

A two-point correlation function is related to the power spectrum by a
Fourier transform,
\begin{align}
\xiXs(\bfr)&=\int\frac{d^3\bfk}{(2\pi)^3}\PXs(\bfk)\,e^{i\,\bfk\cdot\bfr} \nn \\
&=\int\frac{d^3\bfk}{(2\pi)^3}\PhXs(\bfk)\Dfog^2(fk\mu_{\bfk}\sigv)\,e^{i\,\bfk\cdot\bfr}.  \label{eq:fourier}
\end{align}
We start by considering the power spectra of the shape field, namely
Eqs.~(\ref{eq:Pgp_general})--(\ref{eq:Ppm_general}). Following the
model developed in Sec.~\ref{sec:fog}, the part $\PhXs(\bfk)$ in
Eq.~(\ref{eq:redshift_power}) are given by
\begin{align}
& \Phgps (\bfk)  = b_Kk^{-2}(k_x^2 - k_y^2)\left[b \Pdd(k)+f\mu_{\bfk}^2\Pdt(k)\right], \label{eq:pk_gi}\\
& \Phpms(\bfk)  = b_K^2 k^{-4}\left[ (k_x^2 - k_y^2)^2  \pm ( 2 k_x k_y)^2\right] \Pdd(k), \label{eq:pk_iip} \\ &\Phgms(\bfk) = \Phppmms(\bfk) = \Phmmpps(\bfk) =0.
\end{align}
In the following, we derive the multipoles of the non-zero IA power
spectra, $\Pgps(\bfk)$ and $\Ppms(\bfk)$, and substitute them into
Eq.~(\ref{eq:fourier}). Following Ref.~\cite{Kurita:2022}, we express
the inverse Hankel transform, as
\be
\mathcal{H}_\ell^{-1}\left[ g(k)\right](r) \equiv i^\ell \int \frac{k^2 dk}{2\pi^2}j_\ell(kr)g(k). \label{eq:hankel}
\ee
Using this notation, the commonly-used multipole expansion of an anisotropic power spectrum is described as 
\begin{align}
\xiXs(\bfr)
= \sum_{\ell}\xiXls(r){\cal L}_{\ell}(\mu_s),
\end{align}
where 
$\xiXls(r) = \mathcal{H}_\ell^{-1} \left[ \PXls(k)\right]$.
The multipole components of the conventional galaxy density
correlation function with the FoG Gaussian damping term
(Eq.~\ref{eq:Pggls}) are given by
\begin{align}
\xiggls(r) &= b^2\Pi_{\delta\delta,\ell}^{gg(0)}(r) +2bf \Pi_{\delta\Theta,\ell}^{gg(1)}(r) 
\nn \\ & \qquad\qquad \qquad
+ f^2\Pi_{\Theta\Theta,\ell}^{gg(2)}(r),
\end{align}
where $\Pi_{X,\ell}^{Y(n)}(r) = \mathcal{H}_\ell^{-1} \left[
  \mathcal{Q}_{Y,\ell}^{(n)}(k)P_X(k)\right]$.  In the linear theory
limit ($\sigv \to 0$), the equation is reduced to the Kaiser formula
\citep{Kaiser:1987,Hamilton:1992,Okumura:2011}.

The II($+$) power spectrum, $\Pps(\bfk)$, is equivalent to the EE
power spectrum in our model, $\Pps(\bfk)=\PEE(\bfk)$
[Eqs.~(\ref{eq:PEEls}) -- (\ref{eq:QEE4})].  The expression for the
II($+$) spectrum is expanded in terms of the standard Legendre
polynomials,
$
\Ppls(k)=\PEEls(k)
$,
where $\ell=0,2$ and $4$ multipoles contain the linear contributions.
The II($+$) multipoles are given in a similar manner to the GG multipoles,
\begin{align}
\xipls(r)&=\mathcal{H}_\ell^{-1} \left[ \PEEls(k)\right](r) \nn \\
&=b_K^2\mathcal{H}_\ell^{-1} \left[ \mathcal{Q}_{+,\ell}(k)\Pdd (k)\right]
, \label{eq:xipls}
\end{align}
where taking $\sigv\to 0$ limit again leads to the linear theory
formula of Ref.~\cite{Okumura:2020}.

Next, let us derive the expressions of the GI and II($-$) correlation
functions, $\xigps(\bfr)$ and $\xims(\bfr)$, respectively. Unlike the
II($+$) correlation function, they are naturally expanded by the
associated Legendre polynomials with $m=2$ and $m=4$,
respectively. Using the normalized associated Legendre polynomials,
they are expressed as
\begin{align}
\xigps(\bfr) &= \sum_{L\geq 2}\xitgpls(r)\Theta_{L}^{m=2}(\mu_{\bfr})\cos(2\phi_{\bfr}), \label{eq:GI_xi_multipole} \\
\xims(\bfr) &= \sum_{L\geq 4}\xitmls(r)\Theta_{L}^{m=4}(\mu_{\bfr})\cos(4\phi_{\bfr}), \label{eq:IIm_xi_multipole}
\end{align}
where $\phi_{\bfr}$ is the azimuthal angle of the separation vector
projected along the line-of-sight direction ($z$-axis) and
$\mu_{\bfr}=r_z/r$.  Note that the definitions of $\xigps$ and $\xims$
in this paper are different from those in Ref.~\cite{Kurita:2022} by
the factors of the azimuthal angle (see their Eqs.~(25) and (27)).
%
To predict the multipoles of the GI and II($-$) correlation functions measured from simulations or observations, we make them coordinate invariant by setting $\phi_{\bfr}=0$, as 
\begin{align}
\xitgpls (r) &= \int_{-1}^1 d\mu_{\bfr} \,\Theta_{L}^{m=2}(\mu_{\bfr})\,\xigps (\bfr)\Bigr|_{\phi_{\bfr}=0}
\nn \\ &
=\mathcal{H}_L^{-1}\left[\PtgEls(k) \right](r),
\label{eq:xigp_multipole_associated_legendre} \\
\xitmls (r) &= \int_{-1}^1 d\mu_{\bfr} \,\Theta_{L}^{m=4}(\mu_{\bfr})\,\xims (\bfr)\Bigr|_{\phi_{\bfr}=0} \nn \\
&=\mathcal{H}_L^{-1}\left[\PtEEls(k) \right](r).
\label{eq:xim_multipole_associated_legendre}
\end{align}
The correlation functions with $\phi_{\bfr}=0$ are equivalent with
those defined in Ref.~\cite{Kurita:2022}.  In appendix
\ref{sec:derivation}, we provide alternative ways to derive the above
equations by using the spherical harmonic expansion.

While correlation functions of the projected shape field are naturally
expanded in the associated Legendre basis, those in the standard
Legendre basis is commonly adopted.  In the following we provide the
nonlinear formulas for the multipoles of the IA correlation functions
in redshift space expanded in the standard Legendre basis.

Setting the angle $\phi_{\bfr}$ to zero, the multipole expansion of
$\xigps(\bfr)$ in Eq.~(\ref{eq:GI_xi_multipole}) is analytically given
by
\begin{align}
\xi^{S}_{g+,0}&(r) = -\sqrt{\frac{5}{12}}\,\wt{\xi}_{g+,2}^S(r) + \frac{1}{\sqrt{20}}\,\wt{\xi}_{g+,4}^S(r) 
\nn \\ &
-\frac{1}{2}\sqrt{\frac{13}{210}}
\,\wt{\xi}_{g+,6}^S(r) +
\frac{1}{6}\sqrt{\frac{17}{70}}
\,\wt{\xi}_{g+,8}^S(r) 
\nn \\ &
- \frac{1}{6}\sqrt{\frac{7}{55}}\wt{\xi}_{g+,10}^S(r)
+\frac{5}{\sqrt{12012}} \wt{\xi}_{g+,12}^S(r)
+\cdots  
, \label{xigp0s}  \\
\xi^{S}_{g+,2}&(r) = \sqrt{\frac{5}{12}}\,\wt{\xi}_{g+,2}^S(r)+ \sqrt{\frac{5}{4}}\,\wt{\xi}_{g+,4}^S(r)
\nn \\ & 
-\frac{1}{2}\sqrt{\frac{65}{42}}\,\wt{\xi}_{g+,6}^S(r) + \frac{1}{6}\sqrt{\frac{85}{14}}\,\wt{\xi}_{g+,8}^S(r)
\nn \\ &
-\frac{1}{6} \sqrt{\frac{35}{11 }} \wt{\xi}_{g+,10}^S(r)
+\frac{25}{\sqrt{12012}} \wt{\xi}_{g+,12}^S(r)
+\cdots
, \label{xigp2s}\\
\xi^{S}_{g+,4}&(r) = -\frac{3}{\sqrt{5}}\,\wt{\xi}_{g+,4}^S(r) -\frac{3}{2}\sqrt{\frac{39}{70}}\,\wt{\xi}_{g+,6}^S(r) 
\nn \\& 
+\frac{3}{2}\sqrt{\frac{17}{70}}\,\wt{\xi}_{g+,8}^S(r)-3 \sqrt{\frac{7}{220}} \wt{\xi}_{g+,10}^S(r)
\nn \\ &
+15 \sqrt{\frac{3}{4004 }} \wt{\xi}_{g+,12}^S(r)+\cdots, \label{xigp4s}
\end{align}
and that of $\xims(\bfr)$ in Eq.~(\ref{eq:IIm_xi_multipole}) is given by
%
\begin{align}
\xi^{S}_{-,0}(r) = & \sqrt{\frac{7}{20}} \wt{\xi}_{-,4}^S(r)-\frac{1}{5}\sqrt{\frac{13}{7}}\wt{\xi}_{-,6}^S(r)
\nn \\ & 
+\frac{1}{15} \sqrt{\frac{187}{28 }} \wt{\xi}_{-,8}^S(r)-\frac{2}{3}\sqrt{\frac{13}{385}}\wt{\xi}_{-,10}^S(r)
\nn \\ &
+\frac{25}{6 \sqrt{2002 }} \wt{\xi}_{-,12}^S(r)+\cdots \, ,
\label{xim0s} \\
\xi^{S}_{-,2}(r) = &-\sqrt{\frac{5}{7 }} \wt{\xi}_{-,4}^S(r)-\sqrt{\frac{13}{28}}\wt{\xi}_{-,6}^S(r)
\nn \\ &
+\frac{4}{3} \sqrt{\frac{34}{154 }} \wt{\xi}_{-,8}^S(r) -\frac{43}{3}\sqrt{\frac{5}{4004}}\wt{\xi}_{-,10}^S(r)
\nn \\ &
+\frac{5}{3} \sqrt{\frac{11}{182 }} \wt{\xi}_{-,12}^S(r)+\cdots \, ,
\label{xim2s}\\
\xi^{S}_{-,4}(r) = & \frac{3}{\sqrt{140 }} \wt{\xi}_{-,4}^S(r)+\frac{6}{5}\sqrt{\frac{13}{7}}\wt{\xi}_{-,6}^S(r)
\nn \\ &
+\frac{3}{5} \sqrt{\frac{17}{308 }}\wt{\xi}_{-,8}^S(r)-3\sqrt{\frac{11}{455}}\wt{\xi}_{-,10}^S(r)
\nn \\ &
+\frac{45}{2 \sqrt{2002 }} \wt{\xi}_{-,12}^S(r)+\cdots \, .
\label{xim4s}
\end{align}
Each multipole moment of the GI and II($-$) correlation functions
involving the nonlinear RSD damping factor is expressed by infinite
terms in the standard Legendre basis, unlike in the associated
Legendre basis (see Eqs.~(\ref{eq:xigp_multipole_associated_legendre})
and (\ref{eq:xim_multipole_associated_legendre})).  These are the
equations used to extract cosmological information from the IA
statistics of the SDSS galaxies in Ref.~\cite{Okumura:2023}.

\begin{figure*}[t]
\centering
\includegraphics[width=0.89\textwidth]{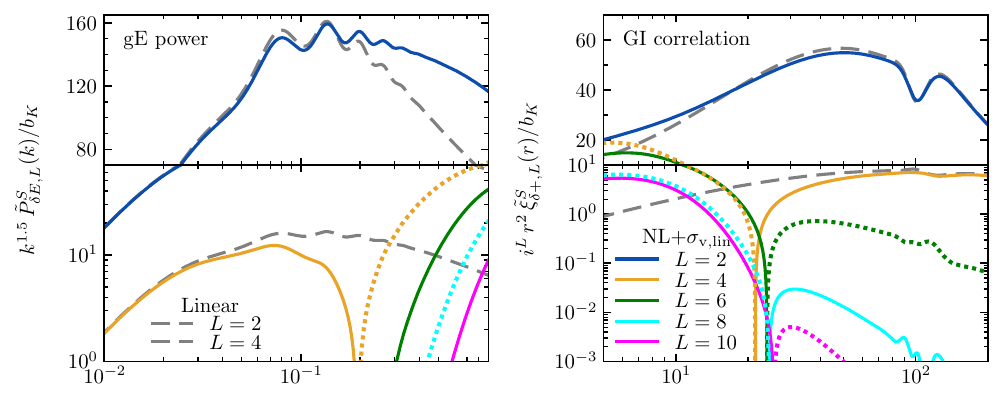}
\includegraphics[width=0.89\textwidth]{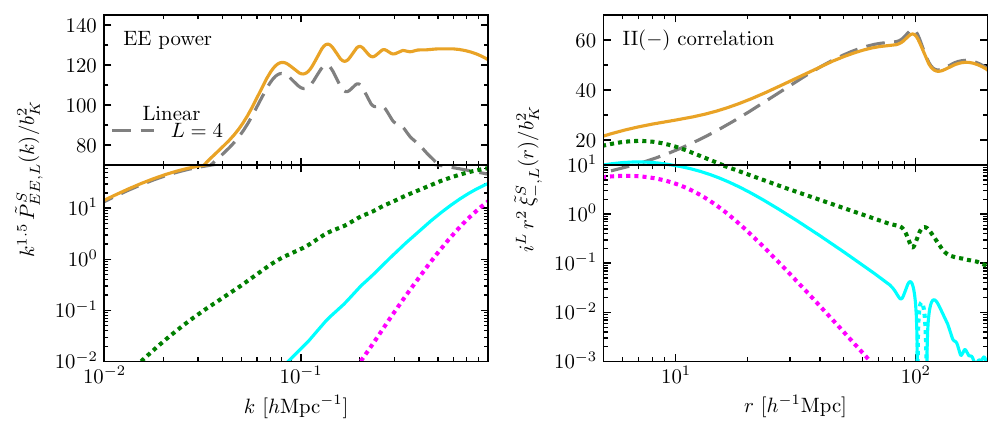}
\caption{ Multipoles of IA statistics expanded in terms of the
  associated Legendre polynomials, the cross power spectra between
  matter density and halo E-mode ellipticity $\PtdEls$ (upper left),
  auto power spectra of halo E-mode ellipticity $\PtEEls$ (lower
  left), cross correlation functions between matter density and halo
  ellipticity $\xitdpls$ (upper right), and auto correlation functions
  of halo ellipticity $\ximls$ (lower left).  The prediction of linear
  theory is adopted for the FoG damping parameter $\sigv$.  The solid
  and dotted curves represent positive and negative values,
  respectively.  The linear theory predictions are shown by the dashed
  gray curves.  }
\label{fig:pkxi_lm_theory}
\end{figure*}

\begin{figure*}[t]
\centering
\includegraphics[width=0.450\textwidth]{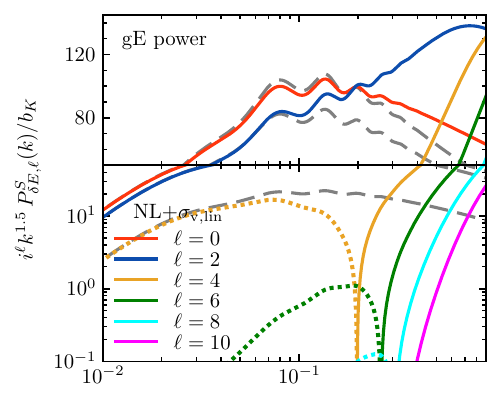}
\includegraphics[width=0.433\textwidth]{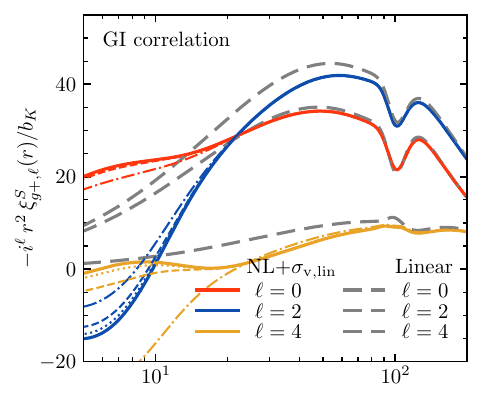}
\includegraphics[width=0.450\textwidth]{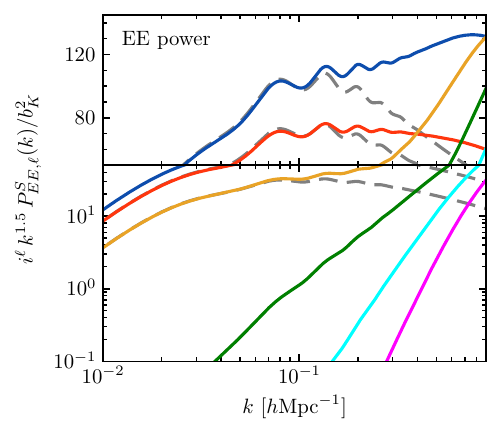}
\includegraphics[width=0.433\textwidth]{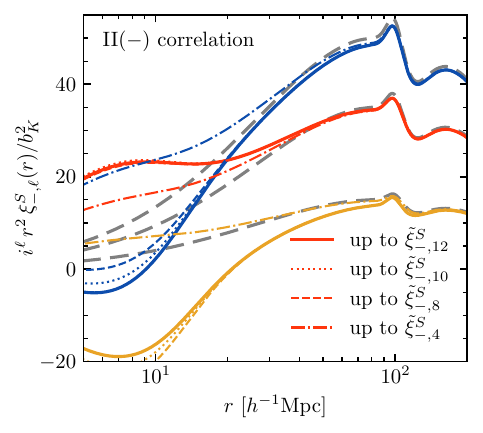}
\caption{Similar with Fig.~\ref{fig:pkxi_lm_theory}, but
Multipoles of IA statistics expanded in terms of the standard Legendre
polynomials.  Since the models of the GI and II($-$) correlation
functions of this basis are expressed by the infinite sums of those of
the associated Legendre polynomials, the upper-right and lower-right
panels show the results of the convergence test, the summation up to
$L=4$, 8, 10 and 12 shown by the dot-dashed, short-dashed, dotted and
solid curves, respectively.  }
\label{fig:pkxi_l_theory}
\end{figure*}

\section{Numerical results}\label{sec:analytic_result}

In this section, we present numerical results of the phenomenological
model of the redshift-space IA statistics derived in the previous
sections.  We present the matter density--halo shape
cross-correlations ($\PtdEls$ and $\xitdpls$) and halo shape
auto-correlations ($\PtEEls$ and $\xitmls$) computed in the associated
Legendre basis, and these in the standard Legendre basis.  Since we
consider for the cross correlations the matter density field, not the
biased object field, we use the symbols $\PtdEls$ and $\xitdpls$,
rather than $\PtgEls$ and $\xitgpls$, respectively (they are
equivalent if we set $b=1$).  We use the publicly-available \class
code \cite{Blas:2011} to compute the linear matter power spectrum
$\Pdd(k)$, and adopt the revised \texttt{Halofit} model to obtain the
nonlinear correction \citep{Takahashi:2012}.  We then use the fitting
formulae derived by \cite{Hahn:2015} to obtain $\Pdt(k)$ and
$\Ptt(k)$.

The function $\Dfog$ is a damping function due to the nonlinear RSD
effect characterized by the one-dimensional velocity dispersion,
$\sigv$. We use the linear theory prediction for $\sigv$, as
\begin{align}
\sigvlin^2=\frac{1}{3}\int \frac{d^3\mathbf{q}}{(2\pi)^3}\frac{\Ptt(q)}{q^2}. \label{eq:sigma_v}
\end{align}

The density--E-mode cross-power spectra $\PtdEls(k)$ and E-mode
auto-power spectra $\PtEEls(k)$ computed in the associated Legendre
basis, are shown in the upper-left and lower-left panels of
Fig.~\ref{fig:pkxi_lm_theory}, respectively.  Since they are
respectively scaled by $b_K$ and $b_K^2$, only the free parameter for
these statistics is $\sigv$ for which we adopt the linear theory
prediction, $\sigvlin$ (Eq.~\ref{eq:sigma_v}).
For $\PtdEls$, only the $L=2$ and $L=4$ spectra have linear-order
contributions, as shown by the gray dashed curves where we set
$\sigv=0$ and $\Pdd=\Pdt=\Ptt$.  The effect of the nonlinear RSD
damping appears prominently in the $L=4$ multipole.  Since the
multipoles with $L\geq 6$ do not contain the linear-order
contributions, they become non-zero only at small and hence nonlinear
scales.
The FoG effect does not have significant contributions to the E-mode
auto power, $\PtEEls(k)$, than to the cross power.

The corresponding configuration-space statistics, GI and II($-$)
correlation functions, are respectively shown in the upper-right and
lower-right panels of Fig.~\ref{fig:pkxi_lm_theory}.  The overall
trend is the same as the case for the power spectra.  Once again, the
nonlinear RSD effect does not impact the quadrupole moment but the
hexadecapole in the associated Legendre basis,
$\wt{\xi}_{\delta+,4}^S$.

The multipoles of the E-mode cross- and auto-power spectra expanded in
the standard Legendre basis, $\PdEls(k)$ and $\PEEls(k)$, are
respectively shown in the upper-left and lower-left panels of
Fig.~\ref{fig:pkxi_l_theory}.  Unlike $\PtdEls$, the nonlinear RSD
effect contributes significantly to not only the hexadecapole but also
the quadrupole for $\PdEls$.

The GI and II($-$) correlation functions expanded in the standard
Legendre basis are shown in the upper-right and lower-right panels of
Fig.~\ref{fig:pkxi_l_theory}, respectively.  Unlike the other
statistics we discussed above, the nonlinear RSD model of the GI and
II($-$) correlation functions expanded by the standard Legendre
polynomials have infinite terms as we saw in Sec.~\ref{sec:nl_2pcf}.
The figure demonstrates that adding the terms up to $\xitXls$ with
$L=12$ makes the multipoles $\xiXls$ ($\ell=0,2,4$) converged even at
small scales of interest.  As is the case with the power spectra,
while the nonlinear RSD effect in the GI correlation functions is
prominent in the hexadecapole in the associated Legendre basis, it is
in the quadrupole in the standard Legendre basis.  However, the
nonlinear RSD effect contributes to the gE power spectrum and GI
correlation function quite differently in the standard Legendre basis.

\section{Comparison to $N$-body simulations}\label{sec:result}
\subsection{$N$-body simulations and Subhalo catalogs}\label{sec:simulation}

As in a series of our papers
\citep{Okumura:2017a,Okumura:2018,Okumura:2019,Okumura:2020a}, we use
$N$-body simulations run as a part of the \darkquest project
\citep{Nishimichi:2019}. We employ $n_p=2048^3$ particles of mass
$m_p= 8.15875\times 10^{10}\himsun$ in a cubic box of side $L_{\rm
  box} = 2\higpc$. In total, we have the data set from eight
independent realizations and we specifically analyze the snapshots at
$z=0.306$.

Halos are identified using the \rockstar algorithm
\citep{Behroozi:2013}.  Their velocities and positions are determined
by the average of the member particles within the innermost 10\% of
the subhalo radius (see Ref.~\citep{Behroozi:2013} for detail).  The
halo mass is defined by a sphere with a radius $R_{\rm 200m}$ within
which the enclosed average density is 200 times the mean matter
density, as $M_h\equiv M_{\rm 200 m}$. We create two halo catalogs,
one with $M_{h} \geq 10^{13}h^{-1}M_\odot$ and another with $M_{h}
\geq 10^{14}h^{-1}M_\odot$, referred to as ``groups'' and
``clusters''.  Note that we remove subhalos, whose center is included
within the sphere of $R_{\rm 200m}$ of a more massive neighbor, from
these samples.
To see the effect of the satellite galaxies on the IA statistics, we
also create mock galaxy catalogs using a halo occupation distribution
(HOD) model \cite{Zheng:2005} applied for the LOWZ galaxy sample of
the SDSS-III Baryon Oscillation Spectroscopic Survey obtained by
Ref.~\cite{Parejko:2013}. We populate (sub)halos with galaxies
according to the best-fitting HOD $N(M_h)$.  After assigning a central
galaxy at the center of a host halo, we randomly draw $N(M_h) - 1$
member subhalos within its $R_{\rm 200m}$ to mimic the positions and
velocities of the satellites. We use a random selection of subhalos
rather than the largest subhalos because a satellite subhalo undergoes
tidal disruption in the host halo and its mass decreases as it goes
toward the center of the gravitational potential.  We call this
subhalo catalog ``HOD luminous red galaxies (LRGs)''.  Properties of
the three subhalo samples constructed above are summarized in Table
\ref{tab:halo}.

Due to the limited hard disk space, the information of dark matter
particles could have been stored partially and thus was lost for four
realizations out of eight. Hence, we could not measure some of the
statistics for which the information of dark matter particles is
needed, while the information of the halos including the direction of
the major axis traced by the dark matter particles was available for
all eight realizations. Thus, when the presented statistics include
the density field of dark matter in the following analysis, the result
is obtained from four realizations; otherwise it is out of the entire
eight realizations.

We assume subhalos to have triaxial shapes \citep{Jing:2002a} and
estimate the orientations of their major axes using the second moments
of the distribution of member particles projected onto the celestial
plane.  The two-component ellipticity of galaxies is defined as
\be
\gamma_{(+,\times)}(\bfx) = \frac{1-q^2}{1+q^2} \left( \cos{(2\phi_x)},\sin{(2\phi_x)} \right), \label{eq:gamma_px}
\ee
where $\phi_x$ is the position angle of the major axis relative to the
reference axis, defined on the plane normal to the line-of-sight
direction, and $q$ is the minor-to-major axis ratio of a galaxy
shape. We set $q$ to zero for simplicity, which corresponds to the
assumption that a galaxy shape is a line along its major axis
\cite{Okumura:2009,Okumura:2009a}.


\begin{table}[tb]
\caption{Properties of mock subhalo samples at $z=0.306$. The quantity
  $f_{\rm sat}$ is the number fraction of satellite subhalos,
  $M_\mathrm{min}$ and $\overline{M}$ are the minimum and average halo
  mass, respectively, $\overline{n}$ is the number density, and $b$
  and $b_K$ are respectively the halo density and shape biases
  computed in the large-scale limit.}
\begin{center}
\begin{tabular}{lcccccc}
\hline\hline
\vspace{-1mm}
\multirow{2}{*}{Types}  &\multirow{2}{*}{$f_{\rm sat}$} & $10^{-12}M_{\rm min}$ &   $10^{4}\overline{n}$ &\multirow{2}{*}{$b$} & \multirow{2}{*}{$b_K$} & $10^{-12}\overline{M}$ \\ 
           &                & $[h^{-1}M_\odot]$ &   $[h^3{\rm ~Mpc}^{-3}]$ &  &  & $[h^{-1}M_\odot]$ \\ 
\noalign{\hrule height 1pt}
Groups  &0 &10 & $ 4.34 $ & $1.66$ & $0.528$ & 32.2 \\ 
Clusters  &0 &100 & $ 0.205 $ & $3.24$ & 0.839 & 188 \\ \vspace{-1.5mm}
HOD &\multirow{2}{*}{0.137} & \multirow{2}{*}{1.63} &   \multirow{2}{*}{$5.27$} &\multirow{2}{*}{$1.72$} & \multirow{2}{*}{0.453} & \multirow{2}{*}{25.2} \\ 
LRGs   & &  &   & &  &  \\
\hline\hline
\end{tabular}
\end{center}
\label{tab:halo}
\end{table}


\subsection{Estimators}

Here, we present estimators to measure from $N$-body simulations the
power spectra and correlation functions of intrinsic halo shapes in
redshift space.

\subsubsection{Power spectra}

Multipole moments of the three-dimensional power spectra of the E-mode
field of halo shapes with the matter/halo distribution, $\PgEls(k)$,
and of the auto-power spectra of the E-mode field, $\PEEls(k)$, have
been measured in the Legendre basis from simulations in
Refs.~\cite{Kurita:2021,Shi:2021}.  These measurements have been
extended to the multipole moments expanded in the associated Legendre
basis in Ref.~\cite{Kurita:2022}, $\PtgEls(k)$ and $\PtEEls(k)$,
respectively.

The density and E-mode shape fields are obtained by assigning the
mass/ellipticity elements of subhalos to a $1024^3$ uniform Cartesian
mesh using the Cloud-In-Cell (CIC) scheme.  We then apply the fast
Fourier transform to estimate the density and E-mode auto-power
spectra, $\Pgg(\bfk)$ and $\PEE(\bfk)$, respectively, and their cross
power spectrum $\PgE(\bfk)$.  Note that we employ the interlacing
technique to reduce the aliasing effect in addition to the
deconvolution of the CIC kernel in Fourier space
\citep{Sefusatti:2016}.  The measured auto-power spectra are affected
by the shot noise.  The Poisson distribution is assumed to estimate
the shot noise for $\Pggs$ and it is subtracted from the monopole
moment in the standard Legendre basis, $P_{gg,0}(k)$.  To estimate the
shot noise for the E-mode auto-power spectrum, $\PEEs$, we measure the
B-mode auto-power spectrum, $\PBBs$, and subtract its constant value
at the large-scale limit from $\PEEs$ to take the non-Poisson shot
noise into account \citep{Kurita:2021}.
Note that even the Poisson shot noise is not orthogonal to any
multipole expanded in terms of the associated Legendre polynomials,
unlike the case of the standard Legendre polynomials \big[$\int^1_{-1}
  d\mu \mathcal{L}_\ell(\mu)=2\delta_{0\ell}$\big]. Thus, the B-mode
auto-power spectra in the associated Legendre basis, $\PtBBls$, need
to be subtracted from $\PtEEls$ with $L$ being arbitrary.

\subsubsection{Correlation functions}

Multipole moments of the correlation functions of galaxy/halo shape
fields have been measured by Refs.~\cite{Singh:2016,Okumura:2020a}.
We use estimators for the multipole correlation functions proposed in
Ref.~\cite{Okumura:2020a}, $\xiXls(r)$ ($X=\{g+, +, - \}$), expressed
as
\begin{equation}
    \xiXls(r) = \frac{2\ell+1}{2} \frac{1}{RR(r)} \sum\limits_{j,k|r=|\bfr_{jk}|} W_{X,jk}\mathcal{L}_{\ell}(\mu_{jk}) , \label{eq:xiXls_est}
\end{equation}
where $\bfr_{jk}=\bfs_{k}-\bfs_{j}$ with $\bfs_j$ the redshift-space
position of $j$-th halo,
$\mu_{jk}=\hat{\bfr}_{jk}\cdot\hat{\bfz}$,\footnote{Here
$\hat{\bfz}=\hat{\bfs}_j=\hat{\bfs}_k$ because the plane-parallel
approximation is assumed.}  and $RR$ is the pair counts from the
random distribution, which can be analytically and exactly computed
because we place the periodic boundary condition on the simulation
box. For the GI and II correlation functions, $W_{g+,jk} =
\gamma_+(\bfs_j)$ and $W_{\pm,jk}=\gamma_+(\bfs_j)\gamma_+(\bfs_k)\pm
\gamma_\times(\bfs_j)\gamma_\times(\bfs_k)$, respectively, where
$\gamma_{(+,\times)}$ is redefined relative to the separation vector
$\bfr_{jk}$ projected on the plane perpendicular to the line of sight,
making the estimated correlation functions coordinate-independent.

By analogy with Eq.~(\ref{eq:xiXls_est}), the multipoles of the GI and
II($-$) correlation functions expanded in terms of the associated
Legendre polynomials are estimated as
\begin{align}
\xitXls(r)=& 
\frac{1 }{RR(r)}\sum\limits_{j,k|r=|\bfr_{jk}|} W_{X,jk}\Theta_{L}^{m}(\mu_{jk}) , \label{eq:xitXls_est}
\end{align}
where $X=\{g+,-\}$ and $m=2,4$, respectively.


\subsection{Determining density and shape bias parameters}\label{sec:bias}

The linear density field in redshift space contains $b$ and $f$ as
parameters, while the linear ellipticity field $b_K$.  The nonlinear
RSD induces the Finger-of-God type damping parameter, $\sigv$, to both
of the fields.  We thus have four parameters, $(f,b,b_K,\sigv)$, with
$f$ being a cosmologically-important parameter and the others nuisance
parameters.

Let us determine the bias parameters, $b$ and $b_K$, using the
real-space statistics.  The density bias parameter can be determined
by the matter-halo cross- or halo auto-power spectra,
\be
b(k) = \frac{\Pdg(k)}{\Pdd(k)}, \qquad
b(k) = 
\left[ \frac{\Pgg(k)}{\Pdd(k)} \right]^{\frac{1}{2}}.
\ee
The parameter can be similarly determined by the corresponding
configuration-space statistics,
\be
b(r) = \frac{\xidg(r)}{\xidd(r)}, \qquad
b(r) = 
\left[ \frac{\xigg(r)}{\xidd(r)} \right]^{\frac{1}{2}}.
\ee

The values of $b_K$ for our halo samples had already been determined
in Refs. \cite{Okumura:2019,Okumura:2020a}.  Here we re-measure them
using the correlators of the halo ellipticity field expanded in the
associated Legendre basis.  In Fourier space, using the cross-power
spectra of the matter density and halo E-mode and the auto-power
spectra of the halo E-mode, respectively, the shape bias is measured
as
\begin{align}
b_K(k) &= -\frac{\sqrt{15}}{4}\frac{\wt{P}_{\delta E,2}(k)}{\Pdd(k)}, \\
b_K(k) &= 
-\left [ \frac{\sqrt{35}}{5}\frac{\wt{P}_{EE,4}(k)}{\Pdd(k)}  \right]^{\frac{1}{2}}.
\end{align}
Similarly in configuration space, using the cross- and auto-correlation functions, the shape bias is respectively estimated as
\begin{align}
b_K(r) &= \frac{\sqrt{15}}{4}\frac{\wt{\xi}_{\delta +,2}(r)}{\xi_{\delta\delta,2}(r)}, \\
b_K(r) &= 
-\left [ \frac{3\sqrt{35}}{16}\frac{\wt{\xi}_{-,4}(r)}{\xi_{\delta\delta,4}(r)}  \right]^{\frac{1}{2}},
\end{align}
where unlike the multipoles $\wt{\xi}_{\delta +,2}(r)$ and
$\wt{\xi}_{-,4}(r)$, $\xidd(r)$ is isotropic and $\xiddl(r)$ is the
Hankel transform of $\Pdd(k)$,
$\xiddl(r)=\mathcal{H}_\ell^{-1}[\Pdd(k)](r)$
[Eq.~(\ref{eq:hankel})]. Since $\xiddl(r)$ with $\ell\neq0$ cannot be
measured from simulations directly in configuration space, we compute
them from the nonlinear matter power spectra $\Pdd(k)$.

\begin{figure}[bt]
\centering
\includegraphics[width=0.495\textwidth]{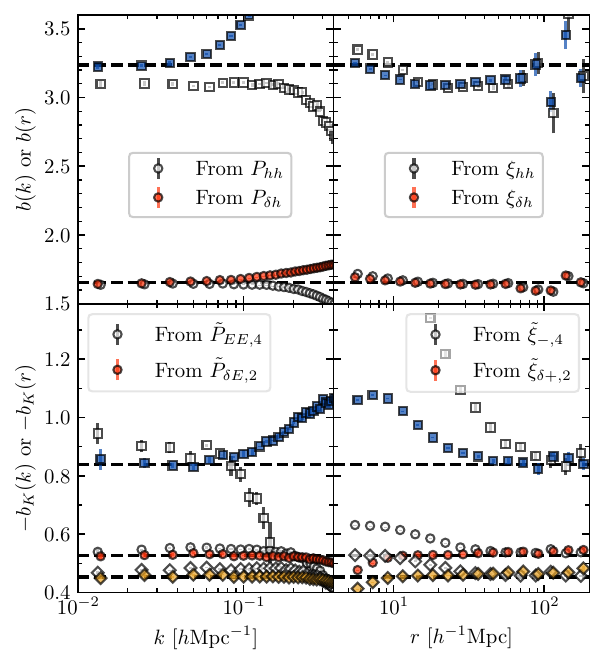}
\caption{ Halo density (upper-left) and shape (lower-left) biases
  determined in Fourier space. The right panes are the same as the
  left ones but the biases determined in configuration space.  The
  full and open points are the biases from the cross correlations with
  the matter field and auto correlation with the halo field,
  respectively.  The blue, red and yellow points are the results for
  clusters, groups and HOD LRGs, respectively. Since the groups and
  HOD LRGs have very similar density bias parameters, we do not show
  the results for HOD LRGs in the upper panels for clarity.  }
\label{fig:bias}
\end{figure}

\begin{figure*}[bt]
\centering
\includegraphics[width=0.9\textwidth]{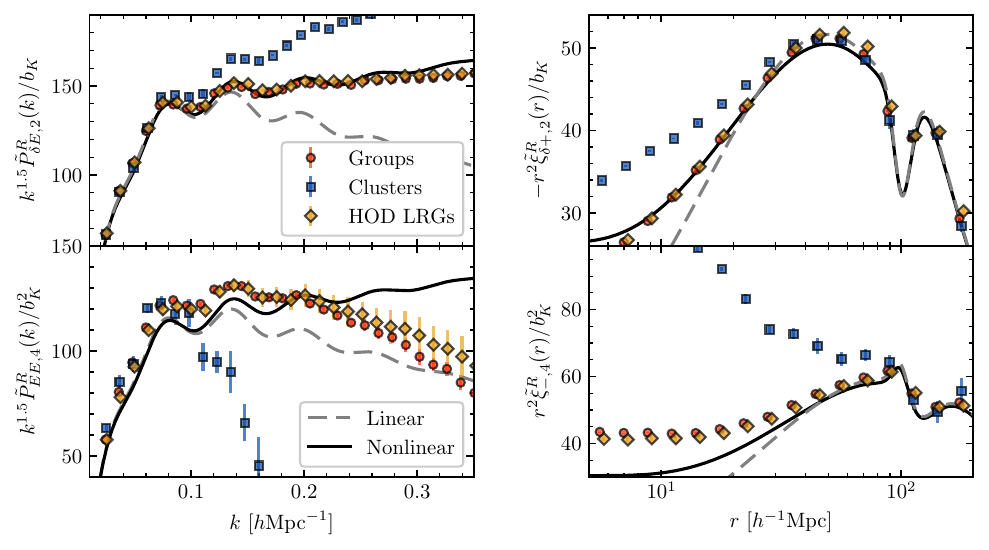}
\caption{ (Left set) IA statistics in real space, the quadrupole
  moment of the matter density - E-mode cross power spectrum (upper
  left), the quadrupole moment of the matter density - shape cross
  correlation function (upper right), the hexadecapole moment of the
  halo E-mode auto power spectrum (lower left), and the hexadecapole
  moment of the halo shape auto correlation function (lower right).
  The solid and dashed curves are the predictions of NLA and LA
  models, respectively.
}
\label{fig:xi_lm_real}
\end{figure*}

\begin{figure*}[bt]
\centering
\includegraphics[width=0.99\textwidth]{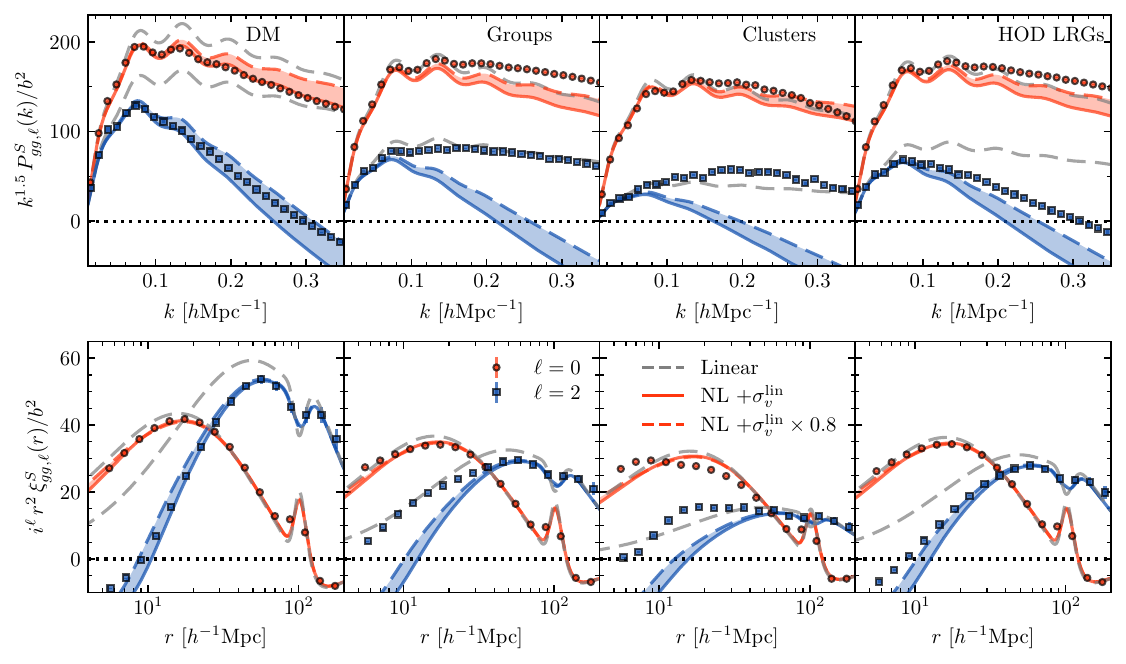}
\caption{ Redshift-space GG power spectra (upper set) and correlation
  functions (lower set).  From the left to right, we show the results
  for dark matter, groups, clusters and HOD LRGs.  The red and blue
  points are the measurements of the monopole and quadrupole moments,
  respectively.  The solid curves are the nonlinear RSD model with the
  velocity dispersion predicted by linear theory, $\sigv^{\rm lin}$.
  The shaded regions indicate the model with the values of $\sigv$ of
  $0.8\times\sigvlin \leq \sigv \leq \sigvlin$.  The dashed gray
  curves are the linear theory prediction.  }
\label{fig:xigg_l_red}
\end{figure*}

\begin{figure*}
\centering
\includegraphics[width=0.95\textwidth]{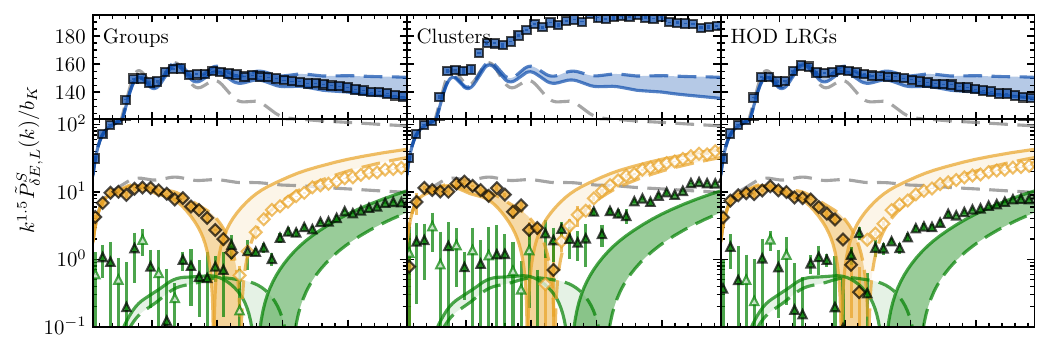}
\includegraphics[width=0.95\textwidth]{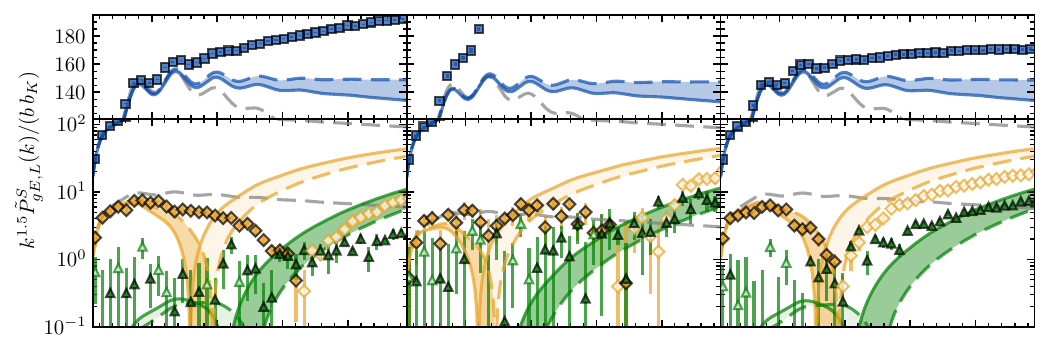}
\includegraphics[width=0.95\textwidth]{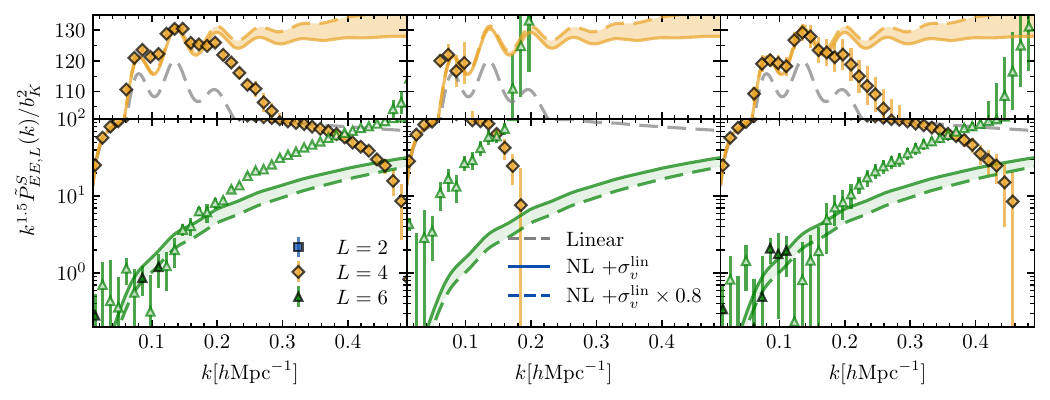}
\caption{ Multipoles of redshift-space IA power spectra expanded in
  terms of associated Legendre polynomials: gE power between matter
  density and halo E-mode ellipticity $\PtdEls$ (first row), gE power
  between halo density and halo E-mode ellipticity $\PtgEls$ (second
  row), and EE power of halo E-mode ellipticity $\PtEEls$ (third row).
The solid curves are the nonlinear RSD model with the velocity
dispersion predicted by linear theory, $\sigvlin$.  The shaded regions
indicate the model with the values of $\sigv$ of $0.8\sigvlin \leq
\sigv \leq \sigvlin$.  Linear theory predictions for the multipoles
$\PtXls$ with $L\leq 4$ are shown by the gray curves.  }
\label{fig:pk_lm_red}
\end{figure*}

\begin{figure*}[t]
\centering
\includegraphics[width=0.95\textwidth]{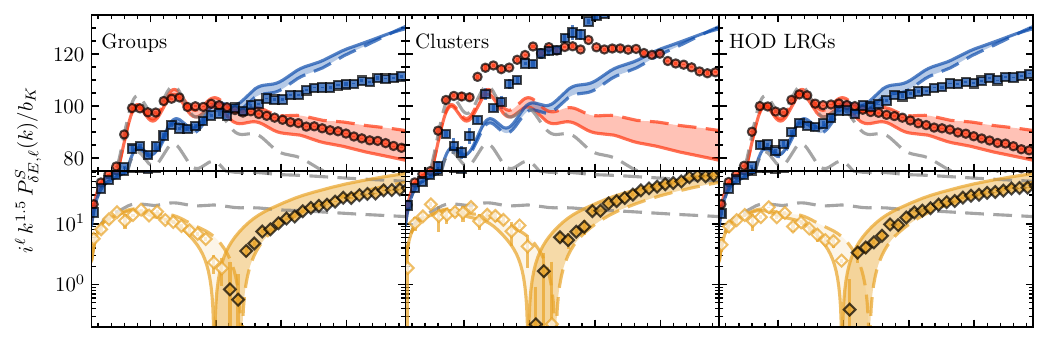}
\includegraphics[width=0.95\textwidth]{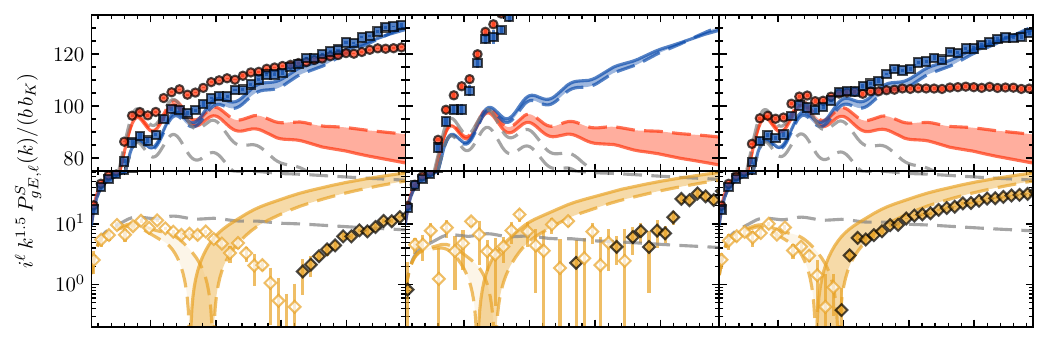}
\includegraphics[width=0.95\textwidth]{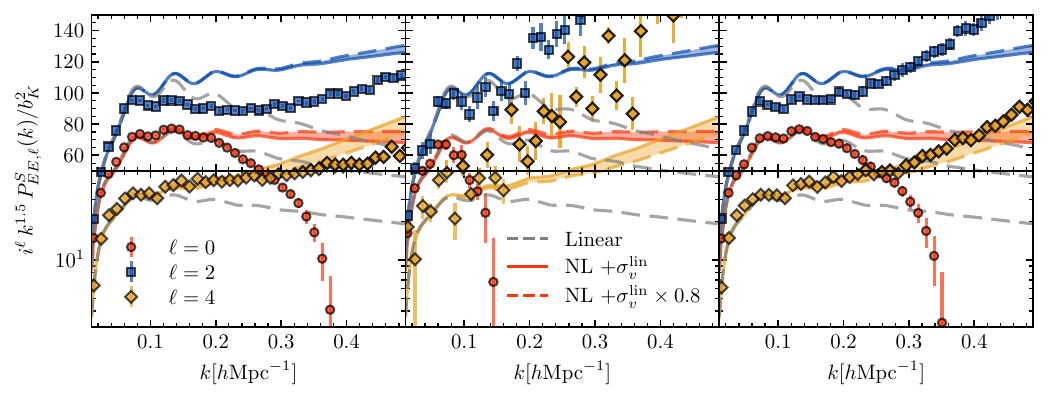}
\caption{ Similar to Fig.~\ref{fig:pk_lm_red} but multipoles of
  redshift-space IA power spectra expanded in terms of standard
  Legendre polynomials: gE power between matter density and halo
  E-mode ellipticity $\PdEls$ (first row), gE power between halo
  density and halo E-mode ellipticity $\PgEls$ (second row), and EE
  power of halo E-mode ellipticity $\PEEls$ (third row).  }
\label{fig:pk_l_red}
\end{figure*}

\begin{figure*}[t]
\centering
\includegraphics[width=0.95\textwidth]{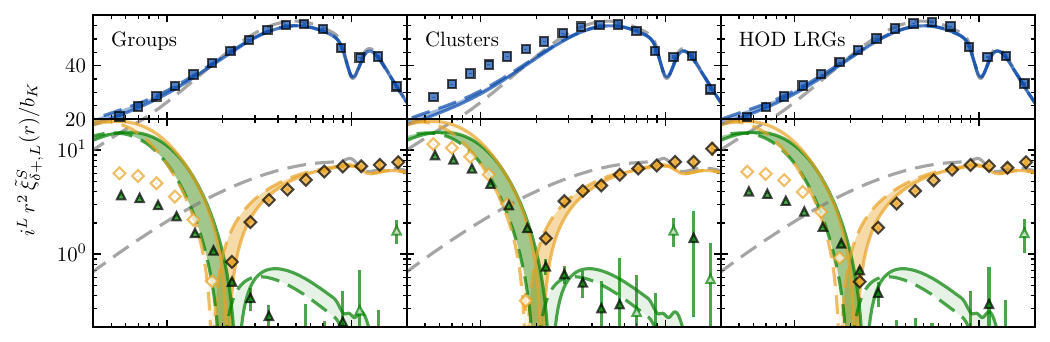}
\includegraphics[width=0.95\textwidth]{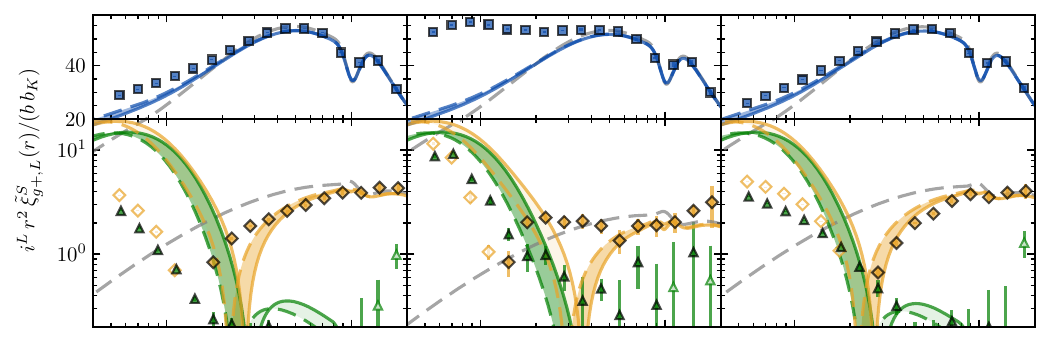}
\includegraphics[width=0.95\textwidth]{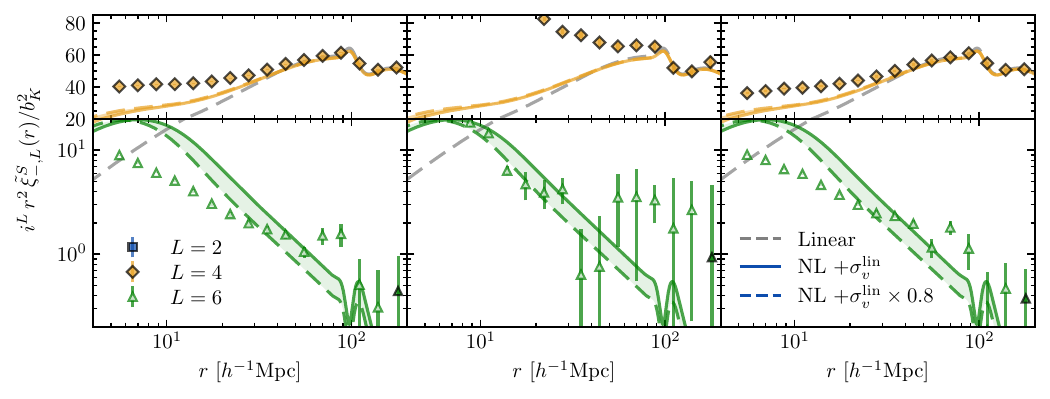}
\caption{ Similar to Fig.~\ref{fig:pk_lm_red} but for multipoles of
  redshift-space IA correlation functions expanded in terms of
  associated Legendre polynomials:
GI correlation between matter density and halo ellipticity $\xitdpls$ (first row),
GI correlation between halo density and ellipticity $\xitgpls$ (second row), and 
II($-$) correlation of halo ellipticity $\xitmls$ (third row).
}
\label{fig:xi_lm_red}
\end{figure*}

\begin{figure*}[t]
\centering
\includegraphics[width=0.95 \textwidth]{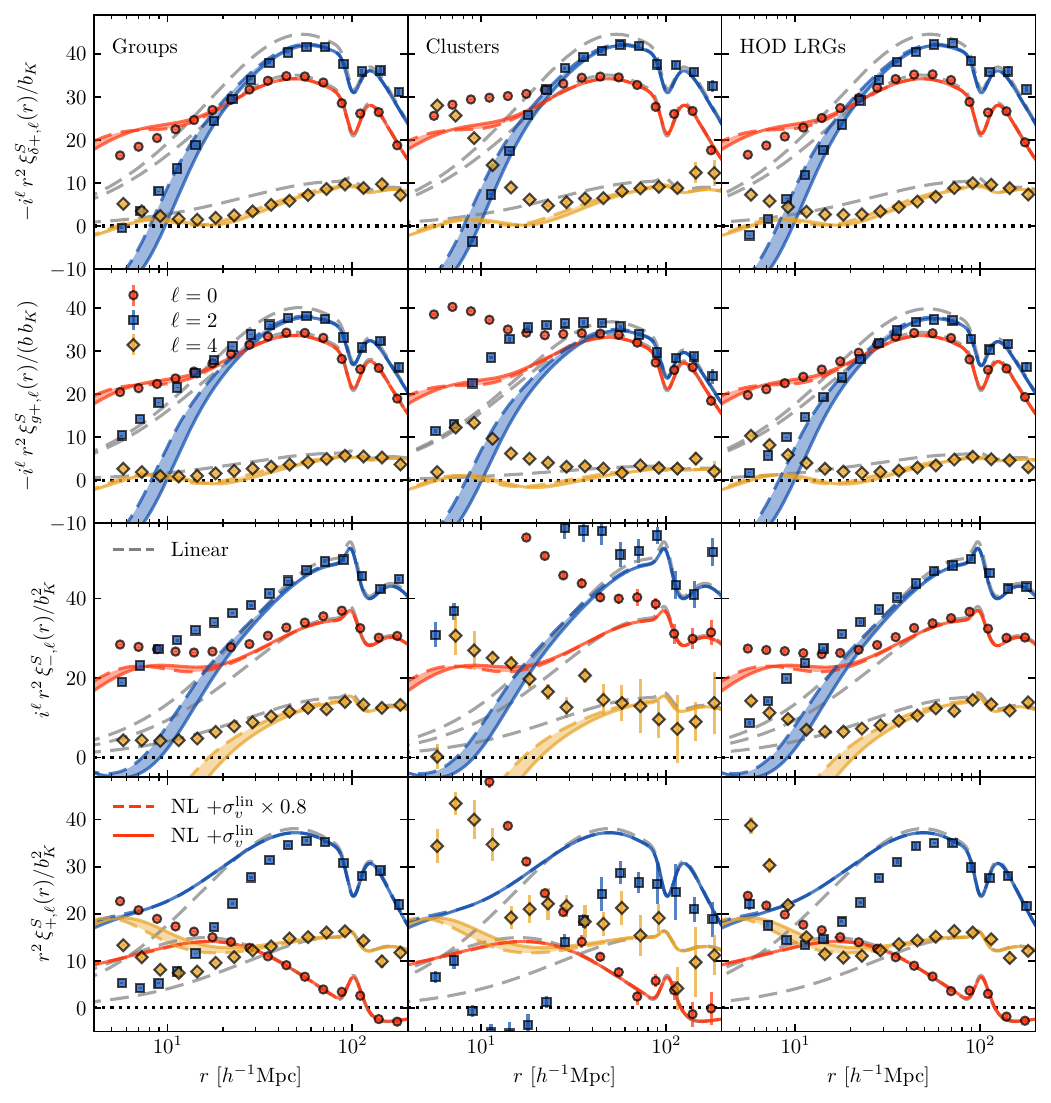}
\caption{
Similar to Fig.~\ref{fig:xi_lm_red} but for multipoles of redshift-space IA correlation functions expanded in terms of standard Legendre polynomials:
GI correlation between matter density and halo ellipticity $\xidpls$ (first row),
GI correlation between halo density and ellipticity $\xigpls$ (second row), and
II($\mp$) correlations of halo ellipticity $\ximpls$ (third and fourth rows).
Our model of the GI and II($-$) correlation functions in the standard Legendre basis contains infinite series of terms, and here we show the modeling results summed up to the twelfth order (see the text).
}
\label{fig:xiia_l_red}
\end{figure*}

Fig.~\ref{fig:bias} shows the halo biases defined above and determined
from simulations.  The density bias parameters, $b$, are shown in the
upper panels.  The shot noise was corrected for the bias determined
from the auto-power spectrum assuming the Poisson distribution.  On
the other hand the bias obtained from the cross-power spectrum tends
to have larger values at high-$k$, particularly for more massive
halos.  The discrepancy is due to the deviation of the shot noise from
the Poisson distribution. Thus the discrepancy is severer for massive
halos and the bias determined from the auto-power spectrum is
suppressed at high-$k$.  They are common features seen in earlier
studies (see e.g., Fig.~2 of Ref.~\cite{Okumura:2012b}).  The density
bias parameters determined from cross- and auto-correlation functions
in configuration space for such massive halos also tend to be
scale-dependent and deviate from the correct values due to the
non-linearity. The matter-halo cross-power spectrum thus provides the
most reliable estimate of the density bias for massive halos and we
use the large-scale limit of the $b(k)$ values from the cross spectrum
to determine the linear bias.  The resultant bias values are shown in
Table \ref{tab:halo}.

The measured shape bias parameters, $b_K$, are shown in the lower
panels of Fig.~\ref{fig:bias}.  The shape bias parameters from auto-
and cross-power spectra, $\wt{P}_{EE,4}$ and $\wt{P}_{\delta E,2}$,
respectively, behave very similarly, except for the massive halos.
The shape field is more severely affected by the non-Poisson shot
noise than the density field \cite{Kurita:2021}, and it cannot be
properly subtracted even though we use the large-scale limit of
$\wt{P}_{BB,4}$.  The shape bias parameters determined from the
cross-power spectra are well consistent with those from both the auto-
and cross-correlation functions, $\wt{\xi}_{+,4}$ and
$\wt{\xi}_{g+,2}$, respectively.  The parameter determined from
$\wt{\xi}_{+,4}$ starts to deviate from the constant at larger scales
than that from $\wt{\xi}_{\delta +,2}$, since the shape field is
density-weighted and thus the shape auto-correlation is more severely
affected by it.  Similarly to the case of the density bias, the linear
shape bias parameter is determined by the large-scale values of
$b_K(k)$ from the cross-power spectrum and shown in Table
\ref{tab:halo}.  Note that, as we set $q = 0$ in
Eq.~(\ref{eq:gamma_px}), the definition of $b_K$ here is different
from literature and one cannot directly compare the values.  It is
interesting to note that the HOD LRG sample has a lower $b_K$ value
than the group sample though they have similar density bias $b$
values.  It is because the existence of satellite galaxies/subhalos
tends to increase the density bias $b$ but decrease the shape bias
$b_K$ due to the misalignment between the major axes of satellites and
their host halos.

The left panels of Fig.~\ref{fig:xi_lm_real} show the cross- and
auto-power spectra of the shape field in real space, $\wt{P}_{\delta
  E,2}$ and $\wt{P}_{EE,4}$, respectively.  The right panels of
Fig.~\ref{fig:xi_lm_real} are similar to the left panels but show the
cross- and auto-correlation functions of the shape field,
$\wt{\xi}_{\delta +,2}$ and $\wt{\xi}_{-,4}$, respectively.  Both in
Fourier space and configuration space, the cross- and
auto-correlations are divided by the best-fitting value of $b_K$ and
its square obtained above, respectively in the figure.
Except for the case of the clusters, both the real-space cross-power
spectra and cross-correlation functions between the matter density and
halo shape fields are well described by NLA model predictions with the
linear shape bias.  Similar results are obtained for the shape
auto-power spectra and auto-correlation functions, but discrepancies
with the model predictions start to appear at larger scales.

\subsection{Model comparison with $N$-body results}

Using the bias parameters, $b$ and $b_K$, determined in the previous
subsection, here we compare our model predictions of galaxy
ellipticity correlations in redshift space with $N$-body measurements.
There is another nuisance parameter, the velocity dispersion parameter
$\sigv$.  The best-fitting parameter of $\sigv$ strongly depends on
RSD models selected as well as the choice of the biased density field
\cite{Nishimichi:2011}: The value slightly larger and smaller than the
linear theory prediction, $\sigvlin$, is preferred for the dark matter
and biased objects, respectively.  The deviation of the best-fitting
value from $\sigvlin$ gets larger for incorrect models of RSD and a
broader fitting range, as indicated by a higher value of $k_{\rm
  max}$. In the following, we thus do not fit the $\sigv$ value with
$N$-body results but rather conservatively show the results for a
range of $0.8\times\sigvlin\leq \sigv\leq\sigvlin$ to indicate the
typical level of theoretical uncertainties due to the FoG effect.
Fig.~\ref{fig:xigg_l_red} shows the comparison of the nonlinear model
predictions for the redshift-space power spectra, $\Pggls$, and
correlation functions, $\xiggls$, with the measurements from $N$-body
simulations.  The redshift-space power spectrum and correlation
function for dark matter are well predicted by the RSD model with the
damping factor with $\sigvlin$.  Since the group and cluster samples
do not contain subhalos, the measurements are consistent with the
linear Kaiser model but not with the nonlinear RSD model with the
damping function, as expected.

Figs.~\ref{fig:pk_lm_red} -- \ref{fig:xiia_l_red} provide comparisons
of our nonlinear RSD model predictions of IA statistics to the
$N$-body results.  In these figures results are shown for the group,
cluster and HOD LRG samples from left to right, respectively.
Figs.~\ref{fig:pk_lm_red} and Figs.~\ref{fig:pk_l_red} show the
results for the power spectra of the halo shape field expanded in
terms of the associated and standard Legendre polynomials,
respectively. Figs.~\ref{fig:xi_lm_red} and \ref{fig:xiia_l_red} are
similar with Figs.~\ref{fig:pk_lm_red} and \ref{fig:xi_lm_red},
respectively, but show the results for the Fourier-counterparts,
correlation functions.  We will discuss in detail the results in the
rest of this subsection.

\subsubsection{IA power spectra}
The first row of Fig.~\ref{fig:pk_lm_red} shows the cross-power
spectra of matter density and halo E-mode fields, $\PtdEls(k)$.  The
ratio $\PtdEls(k)/b_K$ does not depend on the bias parameters, $b$ and
$b_K$. We thus show measured $\PtdEls$ divided by the best-fitting
value of $b_K$ determined in Sec.~\ref{sec:bias}.  The measured
quadrupole moments $\wt{P}_{\delta E,2}^S$, the lowest-order
multipoles, are well predicted for groups and HOD LRGs by our
nonlinear RSD model with the velocity dispersion predicted by linear
theory, $\sigvlin$.  On the other hand, there is a large discrepancy
for clusters at $k>0.1\hmpci$.  Interestingly, our model for the
hexadecapole moment, $\wt{P}_{\delta E,4}^S$, well explains the
measured ones not only for groups and HOD LRGs but also for clusters.
The hexadecapole is severely affected by the nonlinear RSD effect, and
its sign flips at around $k\sim 0.2\hmpci$, the scale depending on the
typical value of $\sigv$, and thus the LA model fails to predict the
measured hexadecapole.  Furthermore, our model provides qualitatively
good agreement with the the fully-nonlinear, higher-order moment,
$\wt{P}_{\delta E,6}^S$, measured for all the shape samples.

The second row of Fig.~\ref{fig:pk_lm_red} shows the cross-power
spectra of halo density and E-mode fields, $\PtgEls$.  While the
overall trend is similar with $\PtdEls$ in the first row, here we see
the extra contribution of the halo density bias.  Since we assume the
simplest linear bias, the discrepancy between the model and
measurement starts to appear at lower $k$, and gets more significant
for more massive halos, as seen in the result for clusters (halos with
masses of $M_h\geq 10^{14}h^{-1}M_\odot$ ).

The third row of Fig.~\ref{fig:pk_lm_red} shows the auto-power spectra
of the halo E-mode field, $\PtEEls$.  While this quantity is not
affected by the halo density bias at linear order, it is by the shot
noise.  We measure the B-mode power spectra in the same basis,
$\PtBBls$, and subtract their large-scale limits from $\PtEEls$.  This
estimation of the shot noise becomes more incorrect for more massive,
thus rarer halos.  Our model therefore fails to predict the
measurements of $\wt{P}_{EE,4}$ at $k>0.1\himpc$ and $\wt{P}_{EE,6}$
at all the scales for clusters.  On the other hand, the model works
reasonably well at $k<0.2\himpc$ for less massive halos, namely groups
and HOD LRGs.

As seen in Fig.~\ref{fig:pk_l_red}, the agreement between the models
and measurements of the IA power spectra expanded in terms of the
standard Legendre polynomials is similar with that in
Fig.~\ref{fig:pk_lm_red}.  It is expected because they are equivalent
quantities but expanded by the different basis.  However, unlike the
EE power spectrum in the associated Legendre basis, $\PtEEls$, only
the monopole of that in the standard Legendre basis, $P_{EE,0}^S$, is
affected by the shot noise and thus suppressed significantly at
high-$k$ due to the non-Poissonian shot noise contribution.  It is
interesting to note that $\PEEls$ ($\ell = 0,2,4$) are noisier than
$\wt{P}_{EE,4}^S$ because the linear information encoded in the latter
is split into the three multipoles in the standard Legendre basis.

\subsubsection{IA correlation functions}

Fig.~\ref{fig:xi_lm_red} shows the results similar to
Fig.~\ref{fig:pk_lm_red} but for the correlation functions.  The
first, second and third rows are respectively multipoles of the GI
correlation for matter density and halo shape fields, $\xitdpls$, GI
correlation for halo density and halo shape fields, $\xitgpls$, and
II($-$) auto-correlation for halo shape field, $\xitmls$, expanded in
the associated Legendre basis.  The comparison of our model
predictions to the $N$-body measurements shows a very similar tendency
with the Fourier-space results: the measurements of $\xitdpls$ are in
good agreement with our models, and the agreement gets worse for
$\xitgpls$, particularly for the cluster shape field.  One exception
is that the auto-correlation in configuration space is not affected by
the shot noise as severely as in Fourier space.  Thus, one can see a
reasonable agreement between the predictions and measurements for
$\xitmls$, and even the hexadecapole of clusters is correctly
predicted at the large-scale limit, unlike $\wt{P}_{EE,4}$.

Unlike the power spectra, the nonlinear redshift-space correlation
functions expanded in terms of the standard Legendre polynomials
behave differently from those of the associated Legendre polynomials,
as shown in Fig.~\ref{fig:xiia_l_red}.  Our nonlinear RSD model of the
GI and II($-$) correlation functions expanded in terms of the standard
Legendre polynomials contains infinite series of the associated
Legendre polynomials. As shown in Fig.~\ref{fig:pkxi_l_theory}, the
model converges by adding the term up to sufficiently higher-order.
We computed the expansion up to the twelfth order and confirmed the
convergence of the formula.  We thus show the modeling results summed
up to the twelfth order.
The first row shows the GI correlation functions between matter
density and halo shape fields, $\xidpls$.  The results for the
quadrupole moments for all the halo samples are well explained by our
nonlinear RSD model.  The second row shows the GI correlation
functions between halo density and shape fields, $\xigpls$.  Agreement
between the predictions and measurements gets worse than the case of
$\xidpls$, due to the nonlinear density bias effect.  The third and
fourth rows show the II($\mp$) correlation functions, $\ximls$ and
$\xipls$, respectively.  The standard Legendre coefficients of the
II($-$) correlations, $\ximls$, are noisier than the associated
Legendre coefficients, $\xitmls$, since for the latter the linear
contribution is compressed to only one, hexadecapole moment
$\wt{\xi}_{-,L}^S$.

The model constructed for the HOD LRG sample is very close to the one
used to constrain the growth rate from the SDSS survey
\cite{Okumura:2023}. In Ref.~\cite{Okumura:2023}, we used the monopole
and quadrupole moments of the GI and II correlation functions at
$r\geq 10\himpc$ for the cosmological analysis. The right column in
Fig.~\ref{fig:xiia_l_red} demonstrates that our formulas were
qualitatively accurate enough for the analysis except for the II
correlation functions, particularly the quadrupole of the II($+$)
function. The II correlation functions measured from the SDSS galaxy
samples were so noisy that their imperfect models would not have
affect the cosmological constraints. If one wants to use larger shape
samples in future galaxy surveys to constrain cosmological models with
precision, the more accurate modeling of the alignment statistics
needs to be developed \cite{Taruya:2024}.

\section{Conclusions} \label{sec:conclusion}

In this paper, we have presented analytic model for nonlinear
correlators of galaxy ellipticities in redshift space.
Adopting a simple Gaussian damping function to describe the nonlinear
RSD effect, known as the Finger-of-God, we have derived formulas for
the multipole moments of the power spectra of galaxy ellipticity field
in redshift space, expanded in not only the associated Legendre basis,
a natural basis for the projected galaxy shape field, but also the
standard Legendre basis, conventionally used in literature.
The model had been derived for the redshift-space galaxy power spectra
by Ref.~\cite{Scoccimarro:2004,Taruya:2010}, and our model for the
intrinsic alignment (IA) statistics have been derived by analogy with
it.
The multipoles of the correlation functions of the galaxy shape field
are expressed simply by a Hankel transform of those of the power
spectra.

We compared our model with the IA statistics for halos and mock
galaxies measured from $N$-body simulations. The measured statistics
were found to be in a better agreement with our nonlinear RSD model
than the existing linear alignment model.  It is the first test for
the accuracy of nonlinear RSD models of the IA, though the model had
already been used to place cosmological constraints using from the
redshift-space correlation functions of the galaxy shape field
measured from the Sloan Digital Sky Survey in
Ref.~\cite{Okumura:2023}.

A series of papers
\cite{Matsubara:2022,Matsubara:2022a,Matsubara:2023} used integrated
perturbation theory and presented a nonlinear model of the tidal field
tensor, which naturally includes nonlinear RSD (see also
Ref.~\cite{Chen:2024}). However, the models had not been tested
against $N$-body simulation measurements.  Other perturbation theory
approaches, such as the TNS model
\cite{Taruya:2010,Nishimichi:2011,Taruya:2013} and distribution
function approach
\cite{Seljak:2011,Okumura:2012,Okumura:2012b,Vlah:2012,Vlah:2013}, can
also be used to model the nonlinear RSD effect of galaxy shape fields.
These modelings will be investigated for various halo samples and
redshifts in simulations in future work.

We presented the formulas of IA statistics in redshift space expanded
in terms of different bases.  While they should be equivalent, the
speed of the convergence at higher-order multipoles would be different
(see Refs.~\cite{Okumura:2012,Okumura:2012b} for different bases for
the multipole redshift-space power spectra).  The investigation of
this effect based on the Fisher-matrix approach will be presented in
our future work.

\begin{acknowledgments}

We thank the referee for the careful reading and suggestions. 
T.~O. acknowledges support from the Ministry of Science and Technology of Taiwan under Grant Nos. MOST 111-2112-M-001-061- and NSTC 112-2112-M-001-034- and the Career Development Award, Academia Sinica (AS-CDA-108-M02) for the period of 2019 to 2023.
This work was supported by MEXT/JSPS KAKENHI Grant Numbers JP20H05861, JP21H01081 (A.~T. and T.~N.), JP19H00677 and JP22K03634 (T.~N.).
\end{acknowledgments}

\appendix

\section{Alternative derivations of GI and II($-$) correlation functions}\label{sec:derivation}

In Sec. \ref{sec:nl_2pcf} we derived the models for the GI and II($-$)
correlation functions with the nonlinear RSD effect in terms of the
associated Legendre polynomials. In this appendix, we provide the
derivations of the same models but using the spherical harmonic
expansion.

We begin by considering the spherical harmonic expansion of the GI and
II($-$) power spectra, $\PXs(\bfk)$, where $X=\{g+, -\}$ respectively,
\begin{align}
\PXs(\bfk) = \sum_{\ell,m} \PXlms(k)\Ylma(\hat{\bfk}). \label{eq:ylm_expansion}
\end{align}
The coefficients of the spherical harmonic expansion, $\PXlms(k)$, are given by
\begin{align}
\PXlms(k) = \int d^2\hat{\Omega}_{\bfk} \PXs(\bfk)\Ylm(\hat{\bfk}). \label{eq:ylm_expansion2}
\end{align}
To compute $\PXlms(k)$ explicitly, we first write the geometric
factors in $\Pgps(\bfk)$ and $\Pms(\bfk)$ due to the projection in
terms of the spherical harmonics, respectively, as
\begin{align}
&
k^{-2}(k_x^2 - k_y^2)= \sqrt{\frac{8\pi}{15}}  \Bigl[Y_{2,2}(\hat{k})+Y_{2,-2}(\hat{k})\Bigr] ~, \\
&
k^{-4}\left[ (k_x^2 - k_y^2)^2 - ( 2 k_x k_y)^2\right] \nn \\ & \qquad \qquad \qquad= \sqrt{\frac{128\pi}{315}}\Bigl[Y_{4,4}(\hat{k})+Y_{4,-4}(\hat{k})\Bigr] ~.
\end{align}
Next, we also express the RSD factor, the integrand of
Eq.~(\ref{eq:pnalpha}), in terms of the spherical harmonics,
\begin{align}
\mu_{\bfk}^{2n} e^{-\alpha \mu_{\bfk}^2} &=  \sum_{q=0}^\infty\,F_{2q}^{(n)}(\alpha)\,Y_{2q,0}(\hat{k}).
\label{eq:Kaiser_FoG_spherical}
\end{align}
Using the orthogonality of the spherical harmonics, the coefficient
$F_q^{(n)}(\alpha)$ is written as
\begin{align}
F_q^{(n)}(\alpha) &=2\pi \int^1_{-1} d\mu_{\bfk}  \, \mu_{\bfk}^{2n} \, e^{-\alpha \mu_{\bfk}^2} Y_{q,0}(\hat{k}).\label{eq:pnqalpha}
\end{align}
It is related to $p^{(n)}(\alpha)$ (Eq.~\ref{eq:pnalpha}) as
$F^{(n)}_0=\sqrt{\pi}p^{(n)}$.

Substituting these equations into Eq.~(\ref{eq:redshift_power}) with
Eqs.~(\ref{eq:pk_gi}) and (\ref{eq:pk_iip}) and then using
Eq.~(\ref{eq:ylm_expansion2}), the functions $\Pgplms(k)$ and
$\Pmlms(k)$ are given by
\begin{align}
\Pgplms(k) & =   b_K\sqrt{\frac{8\pi}{15}}\sum_{q=0}\Bigl[bF_{2q}^{(0)}\Pdd(k)+fF_{2q}^{(1)}\Pdt(k)\Bigr] \nn \\ 
& \quad \times \int  d^2\hat{\Omega}_k 
 \Bigl[Y_{2,2}(\hat{k})+Y_{2,-2}(\hat{k})\Bigr]  
\nn \\  & \qquad \qquad \times 
 Y_{2q,0}(\hat{k})\Ylma(\hat{k}),\label{eq:pk_IIm_spherical} \\
 \Pmlms(k) & = b_K^2 \sqrt{\frac{128\pi}{315}}\sum_{q=0} F_{2q}^{(0)}  \Pdd(k)\nn \\
&\quad \times  \int d^2\hat{\Omega}_k 
\Bigl[Y_{4,4}(\hat{k})+Y_{4,-4}(\hat{k})\Bigr] \nn \\
& \qquad\qquad\times Y_{2q,0}(\hat{k})Y_{\ell m}^*(\hat{k})~, \label{eq:pk_IIm_spherical} 
 \end{align}
Utilizing the Wigner 3-j symbols, these expressions read 
\begin{align}
\Pgplms(k) &= b_K\sum_{n=0} \left[ bF_{2n}^{(0)}\Pdd(k) + fF_{2n}^{(1)}\Pdt(k)\right]\,\nn \\
&\times \sqrt{\frac{2(2\ell+1)(4n+1)}{3}}
\left(
\begin{array}{ccc}
\ell & 2& 2n\\
0 & 0 & 0
\end{array}
\right) \nn \\ 
&\ \times \Biggl[
\left(
\begin{array}{ccc}
\ell & 2& 2n\\
m & 2 & 0
\end{array}
\right)+\left(
\begin{array}{ccc}
\ell & 2& 2n\\
m & -2 & 0
\end{array}
\right)
\Biggr], \label{eq:Pgplm}
\\
 \Pmlms(k) 
&= b_K^2\sum_{n=0} F_{2n}^{(0)}\Pdt(k) \, \nn \\ 
&\times \sqrt{\frac{32(2\ell+1)(4n+1)}{35}}
\left(
\begin{array}{ccc}
\ell & 4& 2n\\
0 & 0 & 0
\end{array}
\right) \nn \\
&\ \times \Biggl[
\left(
\begin{array}{ccc}
\ell & 4& 2n\\
m & 4 & 0
\end{array}
\right)+\left(
\begin{array}{ccc}
\ell &4& 2n\\
m & -4 & 0
\end{array}
\right)
\Biggr],
\label{eq:Pmlm}
\end{align}
where we used the following formula:
\begin{align}
\int d^2\hat{\Omega}_k &Y_{\ell_1m_1}(\hat{k})Y_{\ell_2m_2}(\hat{k})Y_{\ell_3m_3}(\hat{k}) \nn \\
= & 
\sqrt{\frac{(2\ell_1+1)(2\ell_2+1)(2\ell_3+1)}{4\pi}} \nn \\
& \times  \left(
\begin{array}{ccc}
\ell_1 & \ell_2& \ell_3 \\
0 & 0 & 0
\end{array}
\right)
\left(
\begin{array}{ccc}
\ell_1 & \ell_2& \ell_3\\
m_1 & m_2 & m_3
\end{array}
\right)~. \label{eq:ylm_formula}
\end{align}
%
Among the coefficients $\Pgplms$, the only non-vanishing ones are even
multipoles with $m=\pm2$, $P^{g+}_{\ell,2}=P^{g+}_{\ell,-2}$.
Similarly, the non-vanishing coefficients $\Pmlms$ are even multipoles
with $m=\pm4$, $P^-_{\ell,4}=P^-_{\ell,-4}$.  The explicit expressions
of the non-zero coefficients that contain linear information are
respectively given as follows:
\begin{align}
P_{\ell,2}^{g+}(k)
&= bb_K\mathcal{Q}_{g+, \ell}^{(0)}(\alpha)\Pdd(k) 
\nn \\ & \qquad\qquad 
+ fb_K\mathcal{Q}_{g+,\ell}^{(1)}(\alpha)\Pdt(k), \label{eq:Pgplm_2}
\end{align}
where 
\begin{align}
\mathcal{Q}_{g+,2}^{(n)}(\alpha)& = \frac{\sqrt{30}}{105} \Bigl(7\,F_0^{(n)}-2\sqrt{5}\,F_2^{(n)}+F_4^{(n)}\Bigr),\\
\mathcal{Q}_{g+,4}^{(n)}(\alpha)& = \frac{\sqrt{2}}{1001}\Bigl(143\,F_2^{(n)}-78\sqrt{5}\,F_4^{(n)}\nn \\ & \qquad\qquad 
+7\sqrt{65}\,F_6^{(n)}\Bigr),
\end{align}
and 
\begin{align}
P_{\ell,4}^-(k)
= b_K^2\mathcal{Q}_{-,\ell}(\alpha)\Pdd(k), \label{eq:Pmlm_4}
\end{align}
where
\begin{align}
\mathcal{Q}_{-,4}(\alpha)&=\frac{4\sqrt{14}}{255255}  \left(2431 \sqrt{5} F_0^{(0)}-4420 F_2^{(0)}
\right. \nn \\ &\qquad\qquad\quad \left. 
+918 \sqrt{5} F_4^{(0)} -68 \sqrt{65} F_6^{(0)}
\right. \nn \\ &\qquad\qquad\quad \left. 
+7 \sqrt{85} F_8^{(0)}\right). 
\end{align}
We show the coefficients $F_q^{(n)}$ up to $q=8$ below, which are required to compute the power spectra which contain linear information:
\begin{align}
F_0^{(n)}(\alpha) &= A_0p^{(n)}, \label{eq:F_0} \\
F_2^{(n)}(\alpha) &= \frac{A_2}{2} \left[3p^{(n+1)}-p^{(n)}\right],\\
F_4^{(n)}(\alpha) &= \frac{A_4}{8} \left[35p^{(n+2)}-30p^{(n+1)}+3p^{(n)}\right],\\
F_6^{(n)}(\alpha) &= \frac{A_6}{16} \left[231p^{(n+3)}-315p^{(n+2)}
\right.\nn \\ & \quad \quad \left.
+105p^{(n+1)}-5p^{(n)}\right],\\
F_8^{(n)}(\alpha) &= \frac{A_8}{128} \left[6435 p^{(n+4)} -12012p^{(n+3)}
+6930 p^{(n+2)}\right.\nn \\ & \quad \quad \left.
-1260p^{(n+1)}+35p^{(n)}\right], \label{eq:F_8} 
\end{align}
with $A_q = \sqrt{\pi(2q+1)}$.  The coefficients
$\mathcal{Q}_{X,\ell}^{(n)}$ and $F_q^{(n)}$ required to compute the
higher-order multipoles are given in Appendix
\ref{sec:higher_order_GI_IIm}.

Next, using the above equations we derive the formulas of nonlinear GI
and II($-$) correlation functions of the galaxy/halo shape field in
redshift space,$\xigps(\bfr)$ and $\xims(\bfr)$.  By substituting
Eq.~(\ref{eq:ylm_expansion}) into Eq.~(\ref{eq:fourier}), with the
Rayleigh formula,
 $e^{i\bfk\cdot\bfr} =\sum_{\ell,m}4\pi\,i^\ell \,j_\ell(kr)\,Y_{\ell m}(\hat{k})\,Y_{\ell m}^*(\hat{r})$, 
we have
\begin{align}
\xiXs(\bfr) = \sum_{\ell, m}  \xiXlms (r)\,Y_{\ell m}(\hat{r}), \label{eq:def_Xi_ell_m}
\end{align}
where the function $\xiXlms$ is related to $\PXlms$ defined in the previous section via the Hankel transform,
$
\xiXlms(r)= 
\mathcal{H}_\ell^{-1}\left[ \PXlms(k) \right](r).
$
Since the non-vanishing GI and II($-$) multipoles are restricted to
$m=\pm2$ and $m=\pm4$, respectively, and $\Theta^m_\ell(\mu) =
\sqrt{2\pi} Y_{\ell m}(\mu,\phi=0)$, the expansion of the correlation
functions with the normalized associated Legendre polynomials is given
by
\begin{align}
\xitgpls (r) 
&=
\sqrt{\frac{2}{\pi}} \Xi_{L,2}^{g+}(r)
=\sqrt{\frac{2}{\pi}} \mathcal{H}_L^{-1}\left[P_{L,2}^{g+}(k) \right](r), \label{eq:xigp_multipole_associated_legendre_ylm} \\
\xitmls (r) 
&=
\sqrt{\frac{2}{\pi}}\Xi_{L,4}^-(r)
=\sqrt{\frac{2}{\pi}} \mathcal{H}_L^{-1}\left[P_{L,4}^{g+}(k) \right](r).
\label{eq:xim_multipole_associated_legendre_ylm}
\end{align}
These equations are equivalent with the final expressions of the GI
and II($-$) correlation multipoles in terms of the associated Legendre
basis, given by Eqs.~(\ref{eq:xim_multipole_associated_legendre}) and
(\ref{eq:xim_multipole_associated_legendre}) in
Sec.~\ref{sec:nl_2pcf}.

\section{Higher-order multipoles}\label{sec:higher_order}
In this paper, we provided the formulas of IA statistics expanded in
terms of the associated and standard Legendre polynomials and in the
body text we explicitly wrote down the formulas that contain
contributions of linear theory, $L \leq 4$ and $\ell \leq 4$,
respectively. In this appendix, we provide our formulas for the
higher-order multipoles, up to $L = 12$ and $\ell = 12$.

\subsection{gE power spectra}\label{sec:higher_order_gE}
The gE power spectra expanded in terms of the associated Legendre
polynomials, $\PtgEls$, are given in Eq.~(\ref{eq:PtgEls}).  To
compute the nonlinear contributions up to $L=12$, we need to have
$\mathcal{\wt{Q}}_{gE,L}^{(n)}$ for $4<L\leq 12$, which are given as
\begin{widetext}
\begin{align}
\mathcal{\wt{Q}}_{gE,6}^{(n)}(\alpha)  &=  \frac{\sqrt{2730}}{64}  
\left[p^{(n)}(\alpha) -20 p^{(n+1)}(\alpha)
+70p^{(n+2)}(\alpha)-84 p^{(n+3)}(\alpha)
+33p^{(n+4)}(\alpha) \right],\\
\mathcal{\wt{Q}}_{gE,8}^{(n)}(\alpha)  &=\frac{3\sqrt{1190}}{128} 
\left[-p^{(n)}(\alpha)+35 p^{(n+1)}(\alpha)
-210 p^{(n+2)}(\alpha) +462 p^{(n+3)}(\alpha) 
\right. \nn \\ & \qquad \qquad \qquad \left.
-429 p^{(n+4)}(\alpha)+143p^{(n+5)}(\alpha)\right],\\
\mathcal{\wt{Q}}_{gE,10}^{(n)}(\alpha)&=\frac{3\sqrt{385}}{512}
\left[7p^{(n)}(\alpha)-378 p^{(n+1)}(\alpha)
+3465 p^{(n+2)}(\alpha)-12012p^{(n+3)}(\alpha)
\right. \nn \\ & \qquad \qquad \qquad \left.
+19305 p^{(n+4)}(\alpha)-14586p^{(n+5)}(\alpha)
+4199p^{(n+6)}(\alpha)\right],\\
\mathcal{\wt{Q}}_{gE,12}^{(n)}(\alpha)&=\frac{5\sqrt{3003}}{1024}
\left[
-3p^{(n)}(\alpha)+231 p^{(n+1)}(\alpha)
-3003 p^{(n+2)}(\alpha)+15015 p^{(n+3)}(\alpha)
\right. \nn \\ & \qquad \qquad \qquad \left.
-36465 p^{(n+4)}(\alpha)+46189 p^{(n+5)}(\alpha)
-29393 p^{(n+6)}(\alpha)+7429p^{(n+7)}(\alpha)
\right].
\end{align}
Those expanded in terms of the standard Legendre polynomials,
$\PgEls$, are given in Eq.~(\ref{eq:PgEls}).  To obtain the nonlinear
contributions up to $\ell =12$, we need to have
$\mathcal{Q}_{gE,\ell}^{(n)}$ for $4<\ell\leq 12$, which are given as
\begin{align}
\mathcal{Q}_{gE,6}^{(n)}(\alpha) &=
-\frac{13}{32}\left[ 5p^{(n)}(\alpha)-110p^{(n+1)}(\alpha)+420 p^{(n+2)}(\alpha)
-546 p^{(n+3)}(\alpha)+231p^{(n+4)}(\alpha) \right] ,\\
\mathcal{Q}_{gE,8}^{(n)}(\alpha) &=
\frac{17}{256} \left[ 35p^{(n)}(\alpha) - 1295p^{(n+1)}(\alpha) 
+ 8190p^{(n+2)}(\alpha) - 18942p^{(n+3)}(\alpha) 
\right. \nn \\ & \qquad \qquad \left.
+ 18447p^{(n+4)}(\alpha) - 6435p^{(n+5)}(\alpha) \right], \\
\mathcal{Q}_{gE,10}^{(n)}(\alpha) &=
\frac{21}{512}\left[-63p^{(n)}(\alpha) + 3528p^{(n+1)}(\alpha) 
- 33495p^{(n+2)}(\alpha) + 120120p^{(n+3)}(\alpha)
\right. \nn \\ & \qquad \qquad \left.
 - 199485p^{(n+4)}(\alpha) + 155584p^{(n+5)}(\alpha) 
- 46189p^{(n+6)}(\alpha)\right],\\
\mathcal{Q}_{gE,12}^{(n)}(\alpha) &=
\frac{25}{2048}\left[ 231p^{(n)}(\alpha) - 18249p^{(n+1)}(\alpha) 
+ 243243p^{(n+2)}(\alpha) - 1246245p^{(n+3)}(\alpha) 
 \right. \nn \\ & \qquad \qquad \left.
+ 3099525p^{(n+4)}(\alpha) - 4018443p^{(n+5)}(\alpha)
 + 2615977p^{(n+6)}(\alpha) 
- 676039p^{(n+7)}(\alpha) \right].
\end{align}

\subsection{EE power spectra}\label{sec:higher_order_EE}
Similarly to the gE power spectra, the EE spectra expanded in terms of
the associated Legendre polynomials, $\PtEEls$, are given in
Eq.~(\ref{eq:PtEEls}).  To compute the nonlinear contributions up to
$L=12$, we need to have $\mathcal{\wt{Q}}_{EE,L}$ for $4<L\leq 12$,
which are given as
\begin{align}
\mathcal{\wt{Q}}_{EE,6}(\alpha)& =\frac{3\sqrt{91}}{32}
\left[
- p^{(0)}(\alpha)+15  p^{(1)}(\alpha)
-50  p^{(2)}(\alpha)+70 p^{(3)}(\alpha) 
-45  p^{(4)}(\alpha)+11  p^{(5)}(\alpha)
\right],\\
\mathcal{\wt{Q}}_{EE,8}(\alpha)& =\frac{3\sqrt{1309}}{128}
\left[p^{(0)}(\alpha)-30 p^{(1)}(\alpha)+175 p^{(2)}(\alpha)
-420 p^{(3)}(\alpha)+495p^{(4)}(\alpha)
-286p^{(5)}(\alpha)+65 p^{(6)}(\alpha)\right],\\
\mathcal{\wt{Q}}_{EE,10}(\alpha)& =\frac{3\sqrt{5005}}{256}
\left[-p^{(0)}(\alpha)+49 p^{(1)}(\alpha)
-441 p^{(2)}(\alpha)+1617p^{(3)}(\alpha)
-3003 p^{(4)}(\alpha)
\right. \nn \\ & \qquad \qquad \qquad\left.
+3003 p^{(5)}(\alpha)
-1547p^{(6)}(\alpha)+323 p^{(7)}(\alpha)
\right],\\
\mathcal{\wt{Q}}_{EE,12}(\alpha)& =\frac{15\sqrt{1001}}{4096}
\left[
5p^{(0)}(\alpha)-360 p^{(1)}(\alpha)
+4620p^{(2)}(\alpha)-24024 p^{(3)}(\alpha)+64350 p^{(4)}(\alpha)
\right. \nn \\ & \qquad \qquad \qquad \left.
-97240p^{(5)}(\alpha)
+83980 p^{(6)}(\alpha)-38760 p^{(7)}(\alpha)
+7429 p^{(8)}(\alpha)
\right]
.
\end{align}
Those expanded in terms of the standard Legendre polynomials,
$\PEEls$, are given in Eq.~(\ref{eq:PEEls}), and
$\mathcal{Q}_{EE,\ell}$ for $4<\ell\leq 12$ are given as
\begin{align}
\mathcal{Q}_{EE,6}(\alpha) &=\frac{13}{32}\left[
-5p^{(0)}(\alpha)+115p^{(1)}(\alpha)-530p^{(2)}(\alpha)+966p^{(3)}(\alpha)-777 p^{(4)}(\alpha)+231p^{(5)}(\alpha)
\right]
,\\
\mathcal{Q}_{EE,8}(\alpha) &=\frac{17}{256}\left[
35 p^{(0)}(\alpha)-1330 p^{(1)}(\alpha)+9485 p^{(2)}(\alpha)-27132 p^{(3)}(\alpha)+37389 p^{(4)}(\alpha) 
\right. \nn \\ & \qquad \qquad \left.
-24882 p^{(5)}(\alpha)+6435 p^{(6)}(\alpha)
\right]
,\\
\mathcal{Q}_{EE,10}(\alpha) &=\frac{21}{512}\left[
-63 p^{(0)}(\alpha)+3591 p^{(1)}(\alpha)-37023 p^{(2)}(\alpha)+153615 p^{(3)}(\alpha)-319605 p^{(4)}(\alpha)
 \right. \nn \\ & \qquad \qquad \left.
+355069 p^{(5)}(\alpha)-201773 p^{(6)}(\alpha)+46189 p^{(7)}(\alpha)
\right]
,\\
\mathcal{Q}_{EE,12}(\alpha) &=\frac{25}{2048}\left[
231 p^{(0)}(\alpha)-18480 p^{(1)}(\alpha)+261492 p^{(2)}(\alpha)-1489488 p^{(3)}(\alpha)+4345770 p^{(4)}(\alpha)
 \right. \nn \\ & \qquad \qquad \left.
-7117968 p^{(5)}(\alpha)+6634420 p^{(6)}(\alpha)-3292016 p^{(7)}(\alpha)+676039 p^{(8)}(\alpha)
\right].
\end{align}

\subsection{GI and II($-)$ power spectra}\label{sec:higher_order_GI_IIm}
We showed the GI and II($-$) power spectra expanded in terms of the
spherical harmonics, $P_{\ell, 2}^{g+}$ (Eq.~\ref{eq:Pgplm_2}) and
$P_{\ell, 4}^{-}$ (Eq.~\ref{eq:Pmlm_4}), respectively.  To compute the
them with nonlinear contributions up to $\ell =12$, we need to have
the terms $\mathcal{Q}_{g+,\ell}^{(n)}$ and $\mathcal{Q}_{-,\ell}$ for
$4<\ell \leq 12$, which are given as
\begin{align}
\mathcal{Q}_{g+,6}^{(n)}(\alpha)& = \frac{2\sqrt{105}}{36465} \Bigl(85 \sqrt{13} F_4^{(n)}-442 F_6^{(n)}
+11 \sqrt{221} F_8^{(n)}\Bigr),\\
\mathcal{Q}_{g+,8}^{(n)}(\alpha)& = \frac{2\sqrt{1105}}{20995} \Bigl(19\sqrt{7} F_6^{(n)}-2 \sqrt{1547} F_8^{(n)}+\sqrt{975} F_{10}^{(n)}\Bigr),\\
\mathcal{Q}_{g+,10}^{(n)}(\alpha)&=\frac{3\sqrt{110}}{260015}\left(115 \sqrt{119} F_8^{(n)}-1190 \sqrt{3} F_{10}^{(n)}+323 \sqrt{7} F_{12}^{(n)}\right),\\
\mathcal{Q}_{g+,12}^{(n)}(\alpha)&=\frac{\sqrt{286}}{90045}\Bigl(783 F_{10}^{(n)}-290 \sqrt{21} F_{12}^{(n)}+23 \sqrt{609} F_{14}^{(n)} \Bigr),
\end{align}
and
\begin{align}
\mathcal{Q}_{-,6}(\alpha)  & = \frac{4\sqrt{2}}{323323}\Bigl( 323 \sqrt{455} F_2^{(0)}-1292 \sqrt{91} F_4^{(0)}+3458 \sqrt{7} F_6^{(0)}-84 \sqrt{1547} F_8^{(0)}+77 \sqrt{39} F_{10}^{(0)}\Bigr),\\
\mathcal{Q}_{-,8}(\alpha)  & = \frac{4\sqrt{22}}{7436429}\Bigl( 3059 \sqrt{119} F_4^{(0)}-1932 \sqrt{1547} F_6^{(0)}+25806 \sqrt{7} F_8^{(0)}-4004 \sqrt{51} F_{10}^{(0)}+429 \sqrt{119} F_{12}^{(0)} \Bigr),\\
\mathcal{Q}_{-,10}(\alpha)& = \frac{4\sqrt{110}}{48474225}\Bigl( 30015 \sqrt{7} F_6^{(0)}-5220 \sqrt{1547} F_8^{(0)}+32886 \sqrt{39} F_{10}^{(0)}-9860 \sqrt{91} F_{12}^{(0)}+323 \sqrt{2639} F_{14}^{(0)}\Bigr),\\
\mathcal{Q}_{-,12}(\alpha)& = \frac{8}{420756273}\Bigl(24273 \sqrt{17017} F_8^{(0)}-427924 \sqrt{429} F_{10}^{(0)}+300390 \sqrt{1001} F_{12}^{(0)}-27132 \sqrt{29029} F_{14}^{(0)}+52003\sqrt{273} F_{16}^{(0)}\Bigr).
\end{align}
To obtain then, we need to compute $F_q^{(n)}$ with $q\leq 16$. These quantities for $0\leq q \leq 8$ are shown in Eqs.~(\ref{eq:F_0}) -- (\ref{eq:F_8}). These with $8<q\leq 16$ are obtained as
\begin{align}
F_{10}^{(n)}(\alpha) &= \frac{A_{10}}{256} \Bigl(46189 p^{(n+5)} - 109395p^{(n+4)} + 90090p^{(n+3)} - 30030p^{(n+2)} + 3465p^{(n+1)} -63 p^{(n)} \Bigr),\\
F_{12}^{(n)}(\alpha) &= \frac{A_{12}}{1024} \Bigl(676039p^{(n+6)} - 1939938p^{(n+5)} + 2078505p^{(n+4)}  - 1021020p^{(n+3)}  + 225225p^{(n+2)} 
\nn \\ & \qquad \qquad \quad
- 18018p^{(n+1)} +231p^{(n)} \Bigr),\\
F_{14}^{(n)}(\alpha) &= \frac{A_{14}}{2048} \Bigl(5014575p^{(n+7)} - 16900975p^{(n+6)} + 22309287p^{(n+5)}  - 14549535p^{(n+4)} + 4849845p^{(n+3)}
\nn \\ & \qquad \qquad \quad
 - 765765p^{(n+2)}+ 45045p^{(n+1)} -429p^{(n)}  \Bigr). \\
F_{16}^{(n)}(\alpha) &= \frac{A_{16}}{32768} \Bigl(300540195p^{(n+8)} - 1163381400p^{(n+7)}  + 1825305300p^{(n+6)} - 1487285800p^{(n+5)} 
\nn \\ & \qquad \qquad \quad
 + 669278610p^{(n+4)}  - 162954792p^{(n+3)}+ 19399380p^{(n+2)}- 875160p^{(n+1)} +6435p^{(n)} \Bigr).
\end{align}
\end{widetext}
As we showed in section \ref{sec:nl_2pcf}, each multipole of the IA
correlation functions in the standard Legendre basis are expressed by
infinite terms expanded in terms of the associated Legendre
polynomials. Computing the above quantities is necessary to obtain the
converged predictions for the correlation function multipoles in the
standard Legendre basis as shown in the body text.

\section{Linear theory limit}\label{sec:linear_limit}

The model developed in this paper has a form that is a combination of
the nonlinear alignment model (NLA) multiplied by the Gaussian damping
function due to the nonlinear RSD effect.  As introduced in
Section~\ref{sec:fog}, the damping function is given by $D_{\rm
  FoG}(fk\mu_{\bfk}\sigv) =\exp{\left(-\alpha \mu_{\bfk}^2/2\right)}$,
where $\alpha=(f\sigv k)^2$.  We can remove the effect of the
nonlinear RSD by taking the $\sigv \to 0$ limit.  In this appendix we
provide the formulas with this limit, though they were already given
in our previous work \cite{Okumura:2020,Saga:2023}.

Note again that the multipoles in the associated Legendre basis in
this paper are expanded by the normalized associated Legendre function
$\Theta_L^m$ [Eq.~(\ref{eq:PXs_normalized})] and thus denoted by
tilde.

\subsection{Power spectra}

The gE power spectra expanded in terms of the associated Legendre polynomials in the linear theory limit are given by
\begin{align}
\wt{P}_{gE,2}^S(k)& = \frac{4}{\sqrt{15}}b_K\left[b\Pdd(k) + \frac17 f\Pdt(k)\right],\\
\wt{P}_{gE,4}^S(k)& = \frac{8}{21\sqrt{5}}b_K f \Pdt(k).
\end{align}
Those expanded in terms of the standard Legendre polynomials are by 
\begin{align}
P_{gE,0}^S(k) &= \frac23 b_K\left[b\Pdd(k)+\frac15 f \Pdt(k) \right],\\
P_{gE,2}^S(k) &= -\frac23 b_K\left[b\Pdd(k)-\frac17 f \Pdt(k) \right],\\
P_{gE,4}^S(k) & = -\frac{8}{35}b_Kf\Pdt(k),
\end{align}

The EE power spectrum expanded in terms of the associated Legendre polynomials in the linear theory limit is only $\wt{P}_{EE,4}^S$, given by
\begin{align}
\wt{P}_{EE,4}^S(k)& = \frac{16}{3\sqrt{35}}b_K^2  \Pdd(k),
\end{align}
and those in the standard Legendre basis are by
\begin{align}
P_{EE,0}^S(k) &= \frac{8}{15}b_K^2 \Pdd(k),\\
P_{EE,2}^S(k) &= -\frac{16}{21} b_K^2 \Pdd(k),\\
P_{EE,4}^S(k) & = \frac{8}{35}b_K^2 \Pdd(k). 
\end{align}

\subsection{Correlation functions}

The GI correlation functions expanded in terms of the associated
Legendre polynomials $\xitgpls$ in the linear theory limit ($L=2, 4$)
are given by
\begin{align}
\wt{\xi}_{g+,2}^{S}(r)&= -\frac{4}{7}\sqrt{\frac{1}{15}}\,b_K\left[7b\xi_{\delta\delta ,2}(r) +f\xi_{\delta\theta,2}(r)\right]\, ,
\nonumber
\\
 \wt{\xi}_{g+,4}^{S}(r)&= \frac{8}{21}\sqrt{\frac{1}{5}}\,b_Kf\,\xi_{\delta\theta,4}(r),
\end{align}
and similarly, the non-vanishing coefficient for the II($-$)
correlation $\xitmls$ appears only for $L=4$,
\begin{align}
 \wt{\xi}_{-,4}^S(r)&= \frac{16}{3}\sqrt{\frac{1}{35}}\,b_K^2\,\xi_{\delta\delta,4}(r),
\end{align}
where the functions $\xi_{\delta\delta,L}$ and $\xi_{\delta\theta,L}$ are defined by
$\xi_{\delta\delta,L} (r)= \mathcal{H}_L^{-1}[\Pdd(k)](r)$ and 
$\xi_{\delta\theta,L} (r)= \mathcal{H}_L^{-1}[\Pdt(k)](r)$, respectively.
%

Finally, those expressions in terms of the standard Legendre basis
have $\ell = 0,2$, and $4$ components.  The multipoles of the GI
correlation functions are given by
\begin{align}
\xi^{S}_{g+,0}(r) &= -\frac{2}{3}b_Kb\,\xi_{\delta\delta,2}(r) 
\nn \\ &\quad 
+b_Kf\Bigl\{-\frac{2}{21}\xi_{\delta\theta,2}(r)+\frac{4}{105}\xi_{\delta\theta,4}(r)\Bigr\},
\\
\xi^{S}_{g+,2}(r) &= \frac{2}{3}b_Kb\,\xi_{\delta\delta,2}(r) 
\nn \\ &\quad 
+b_Kf\Bigl\{\frac{2}{21}\xi_{\delta\theta,2}(r)+\frac{4}{21}\xi_{\delta\theta,4}(r)\Bigr\},
\\
\xi^{S}_{g+,4}(r) &= -\frac{8}{35}b_Kf\,\xi_{\delta\delta,4}(r).
\end{align}
The II correlation functions in redshift space are equivalent with
those in real space in the linear theory limit \cite{Okumura:2020}.
The multipoles of the II($+$) and II($-$) correlation functions in the
standard Legendre basis are respectively given by
\begin{align}
\xi_{+,0}^S(r) &= \frac{8}{15}b_K^2\,\xi_{\delta\delta,0}(r) ,
\\
\xi_{+,2}^S(r) &= \frac{16}{21}b_K^2\,\xi_{\delta\delta,2}(r) ,
\\
\xi_{+,4}^S(r) &= \frac{8}{35}b_K^2\,\xi_{\delta\delta,4}(r),
\end{align}
and
\begin{align}
\xi_{-,0}^S(r) &= \frac{8}{15}b_K^2\,\xi_{\delta\delta,4}(r) ,
\\
\xi_{-,2}^S(r) &= -\frac{16}{21}b_K^2\,\xi_{\delta\delta,4}(r) ,
\\
\xi_{-,4}^S(r) &= \frac{8}{35}b_K^2\,\xi_{\delta\delta,4}(r).
\end{align}

\bibliography{ms_prd.bbl}

\label{lastpage}
\end{document}